\documentclass[11pt]{article}

\usepackage{newtxtext,newtxmath}
\usepackage{newtxtt}

\usepackage{graphicx}
\usepackage{hyperref}

\usepackage[letterpaper,margin=1in]{geometry}

\renewenvironment{abstract}
	{\quotation}
	{\endquotation}

\date{}

\makeatletter
\renewcommand{\fnum@figure}{\textbf{Figure \thefigure}}
\renewcommand{\fnum@table}{\textbf{Table \thetable}}
\makeatother

\usepackage{scicite}

\usepackage{url}

\usepackage[most]{tcolorbox}
\usepackage{listings}
\usepackage{xcolor}
\usepackage{xspace} 

\tcbset{
  promptbase/.style={
    enhanced,
    breakable,
    listing only,
    listing engine=listings,
    width=\textwidth,
    coltitle=white,
    attach boxed title to top center={yshift=-1.2mm},
    boxed title style={size=small},
    boxrule=0.6pt,
    arc=1mm,
    left=1.5mm,
    right=1.5mm,
    top=2mm,
    bottom=1.5mm,
    before skip=8pt,
    after skip=10pt,
    listing options={
      basicstyle=\ttfamily\footnotesize,
      columns=fullflexible,
      breaklines=true,
      keepspaces=true,
      showstringspaces=false,
      aboveskip=0pt,
      belowskip=0pt
    }
  }
}

\newtcblisting{systemprompt}{
  promptbase,
  colback=blue!2,
  colframe=blue!65!black,
  colbacktitle=blue!65!black,
  title=\textbf{System message},
  title after break=\textbf{System message (continued)}
}

\newtcblisting{userprompt}{
  promptbase,
  colback=green!2,
  colframe=green!45!black,
  colbacktitle=green!45!black,
  title=\textbf{User message},
  title after break=\textbf{User message (continued)}
}

\newtcblisting{modelreply}{
  promptbase,
  colback=black!2,
  colframe=black!60,
  colbacktitle=black!60,
  title=\textbf{Model reply},
  title after break=\textbf{Model reply (continued)}
}

\def\model{GABLE\xspace}

\def\scititle{
    Integrating adaptive human behavior into epidemic models with large language models
}
\title{\bfseries \boldmath \scititle}

\author{
	Yicheng~Mao$^{1}$,
	Haoyang ~Li$^{2}$,
	Rob~Deardon$^{1,3}$,
    Hongru~Du$^{2\ast}$\and
	\small$^{1}$Department of Mathematics and Statistics, University of Calgary, University Drive NW, Calgary, T2N 1N4, Canada\and\
	\small$^{2}$Department of Systems \& Information Engineering, University of Virginia, Charlottesville, VA, USA\and
    \small$^{3}$Faculty of Veterinary Medicine, University of Calgary, University Drive NW, Calgary, T2N 1N4, Canada\and
	\small$^\ast$Corresponding author. Email: hongrudu@virginia.edu\and
}

\begin{document} 

\maketitle

\begin{abstract} 

\noindent Infectious disease transmission is shaped by patterns of human interaction, which adapt as epidemic conditions change. Capturing these context-dependent behaviors remains a fundamental challenge for epidemic models. Here, we recast this challenge by using large language models (LLMs) to represent adaptive human behavior within mechanistic epidemic models. We operationalize this idea through Generative Adaptive Behavioral Layer for Epidemics (GABLE), which adapts LLMs to infer behavioral responses to epidemic and policy conditions and translates them into age-structured contact matrices coupled to a mechanistic epidemic model. Applied to COVID-19 in France, GABLE reproduced responses in population mixing and age-specific contact structures that remained epidemiologically informative. In short-term forecasting, LLM-generated contact matrices outperformed mobility-driven matrices derived from real-world mobility data, with the largest gains at longer horizons. GABLE also extends beyond forecasting to prospective policy evaluation by projecting behavioral and epidemic responses to candidate interventions before implementation. When supplied with subsequently implemented policies, GABLE reproduced epidemic trajectories and generated distinct responses to alternative policy timing and composition. By leveraging LLMs as a flexible behavioral layer, GABLE provides a framework for coupling context-sensitive behavioral generation with epidemic dynamics.
\end{abstract}

\noindent

\newpage
\section*{Introduction}


Epidemics are coupled behavioral--biological systems. Human behavior determines
when, where, and between whom opportunities for transmission arise, while
infectious disease risks and public-health interventions, in turn, reshape those behaviors~\cite{funk2010modelling, fenichel2011adaptive, bedson2021review}. Mathematical and data-driven models seek
to represent this evolving system, anticipate future disease burden, and
support the design and evaluation of public-health interventions
~\cite{bedford2019new}. Their prospective value therefore depends on the joint
representation of disease transmission and adaptive human response
~\cite{bedson2021review}.

Behavior is nevertheless one of the least directly observed components of an
epidemic system. Contact surveys provide direct measurements
of population mixing
~\cite{gimma2022changes,feehan2021quantifying,jarvis2021impact}, but
representative contact studies remain unavailable for many populations~\cite{prem2021projecting}, require substantial data-collection infrastructure~\cite{dan2025surveyfatigue}, and capture behavior only after it has occurred. Digital mobility data provide a more timely but indirect behavioral signal.
Mobility indicators have been used to drive contact rates and reconstruct
transmission across epidemic waves
~\cite{chang2021mobility,nouvellet2021reduction,
didomenico2026natcommun}, but movement does not uniquely determine interpersonal contact or age-specific mixing, which are more directly linked to opportunities for infectious disease transmission. More fundamentally, both approaches rely on realized behavior, limiting their ability to anticipate how populations will respond beyond observed conditions or under interventions that have not yet occurred.

To enable such prospective simulation, epidemic models can instead represent behavioral adaptation endogenously~\cite{hamilton2024incorporating}. At the population level, contact rates have been coupled to incidence,
prevalence, or perceived risk through predefined response functions
~\cite{dobson2023balancing,haw2022optimizing,weitz2020awareness,
mao2026identifying}. At the individual level, economic and game-theoretic
frameworks derive protective behavior from decisions made alongside epidemic
dynamics
~\cite{saad2023dynamics,martcheva2021effects,pangallo2024unequal,
du2025improving,ash2022disease}. These approaches support prospective
simulation, but they obtain this capability by specifying behavioral mechanisms and functional relationships in advance. Real-world responses emerge from
interacting influences, including policy content and duration, perceived risk,
economic and caregiving obligations, activity setting, adherence fatigue, and
imperfect compliance~\cite{funk2015nine, rikani2026resetting}. Epidemic models therefore still lack a general mechanism for translating such heterogeneous and interacting context into transmission-relevant responses without prescribing separate rules for each input and interaction.

Large language models (LLMs) offer a potential way to fill this gap. Their capacity to
integrate heterogeneous information within a common contextual representation
allows policy restrictions, numerical epidemic indicators, individual
circumstances, and activity-specific considerations to be evaluated jointly
~\cite{achiam2023gpt}. A growing body of work has shown that LLMs can
reproduce aspects of human judgment and behavior across experimental and
real-world settings. Despite systematic biases
~\cite{cheung2025large,xu2026comparing,hu2025generative, gao2025take}, LLMs have recovered
population-level patterns in public-opinion surveys
~\cite{argyle2023out, xu2026comparing,qu2024performance}, reproduced established findings
from economic and social-psychological experiments
~\cite{cui2025large,binz2023using,xie2025using,luo2025large, aher2023using}, and predicted
treatment effects in preregistered social-science studies
~\cite{ashokkumar2026large}. Recent studies have also used LLMs to simulate
behavioral responses to public policies during epidemics and reported
realistic population-level response patterns
~\cite{liu2026simulating,li2026theory}. These findings motivate the use of LLMs as context-sensitive behavioral models that translate complex and potentially unseen conditions into structured population responses. The remaining challenge is to connect these responses to disease transmission within a dynamically coupled epidemic system.

\begin{figure}[!h]
\centering
\includegraphics[
  width=\textwidth,
  height=\textheight,
  keepaspectratio
]{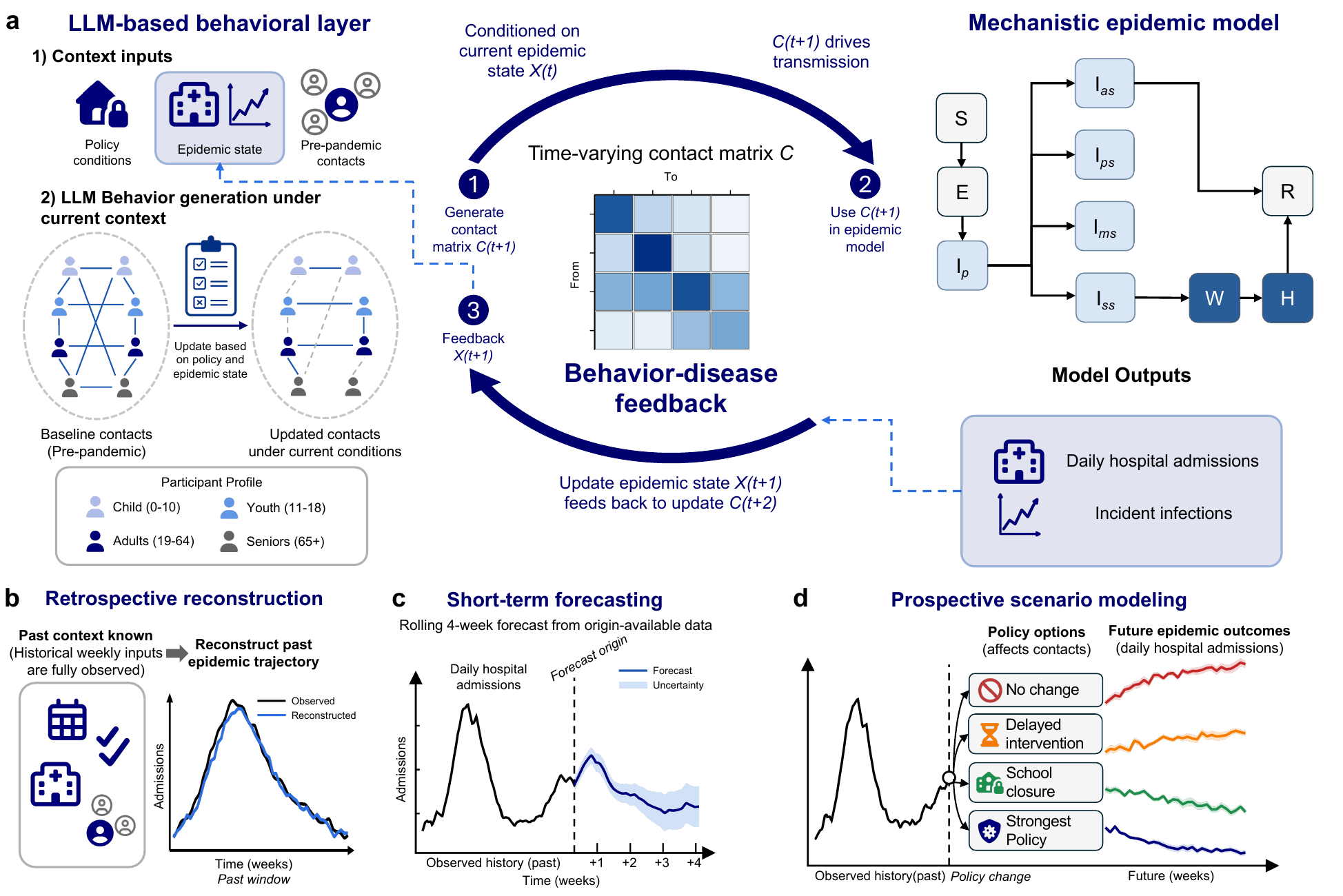}
\caption{\textbf{Overview of \model framework.} (\textbf{a}) The proposed framework couples an LLM-based behavioral layer with a mechanistic compartmental epidemic model through a time-varying contact matrix. The LLM-based behavioral layer takes pre-pandemic contacts with policy and epidemic conditions and uses the LLM to determine how baseline contacts are changed under the current context. These behavioral responses are aggregated into an age-structured contact matrix, which governs the mechanistic epidemic model. The resulting epidemic outcomes are then fed back to the behavioral layer to update subsequent contacts, forming a dynamic behavior–disease feedback loop. (\textbf{b-d}) The proposed framework supports three complementary tasks. (\textbf{b}) Retrospective reconstruction: historical policy and epidemic conditions are used to generate time-varying contacts, and the coupled model is fitted to the observed epidemic trajectory. (\textbf{c}) Short-term forecasting: policies are held fixed and enable rolling forecasts in which future contacts and epidemic outcomes evolve through the coupled behavior–disease system. (\textbf{d}) Prospective scenario modeling: alternative future policies are used to project and compare downstream epidemic outcomes.}
\label{fig:overview}
\end{figure}

Here, we propose the Generative Adaptive Behavioral Layer for Epidemics
(\model), which embeds an LLM-based behavioral layer within a mechanistic
epidemic model through age-structured contact matrices. \model translates changing public-health policies, epidemic
conditions, demographic characteristics, and pre-pandemic contact diaries into
adaptive population mixing patterns, which directly govern transmission in the epidemiological model. During forward simulation, projected epidemic conditions are returned to the behavioral layer, allowing population mixing and disease dynamics to co-evolve. Please see Materials and Methods for full details of the model formulation, implementation, and data inputs. Applied to the COVID-19 epidemic in France, \model generated epidemiologically informative temporal and age-specific changes in population mixing, improved short-term forecasts relative to mobility-driven and statistical reference models, and produced distinct behavioral and epidemic trajectories under alternative future policy scenarios. \model provides a generative behavior–epidemiology framework for prospective epidemic modeling, allowing adaptive human responses and disease dynamics to co-evolve when direct behavioral observations are sparse, delayed, or unavailable.

\section*{Results}
As shown in Fig.~\ref{fig:overview}a, at week $t$, the behavioral layer conditions on the current epidemic state $X(t)$, policy conditions, participant characteristics, and pre-pandemic contacts to generate an updated age-structured contact matrix $C(t+1)$. Details of the LLM inference procedure are provided in the ``LLM prompts and queries'' section in the Supplementary Materials. This matrix governs transmission in an age-structured stochastic epidemic model (see the ``Transmission model'' section in the Supplementary Materials, and fig.~S1), which advances the epidemic to $X(t+1)$; the updated state is then fed back to the behavioral layer to generate $C(t+2)$, forming a sequential behavior-disease feedback loop. We evaluated \model in three settings spanning retrospective reconstruction, short-term forecasting, and prospective policy evaluation; and instantiated the behavioral layer with three LLMs (GPT-4o mini, Gemini 2.5 Flash, and Grok 3 mini). For retrospective epidemic reconstruction (Fig.~\ref{fig:overview}b), we compared the resulting contact matrices with a static pre-pandemic matrix and published mobility-driven matrices \cite{didomenico2026natcommun}, and evaluated their ability to reproduce observed hospitalization dynamics. We propagated all matrices through the same age-structured stochastic transmission model, holding epidemiological structure, parameters, vaccination, and variant inputs fixed. For the short-term forecasting task (Fig.~\ref{fig:overview}c), we conducted rolling four-week forecasts without access to future policy changes or behavioral observations, comparing \model with the mobility-driven mechanistic benchmark and statistical forecasting methods. Finally, for prospective policy evaluation (Fig.~\ref{fig:overview}d), we first tested whether \model reproduced subsequent epidemic trajectories under the policies actually implemented while withholding future hospitalization outcomes, and then applied the same framework to counterfactual interventions that varied policy timing and composition.

\subsection*{LLMs generate context-dependent and age-structured contact adaptation}

\begin{figure}[!h]
\centering
\includegraphics[
  width=0.9\textwidth,
  height=0.67\textheight,
  keepaspectratio
]{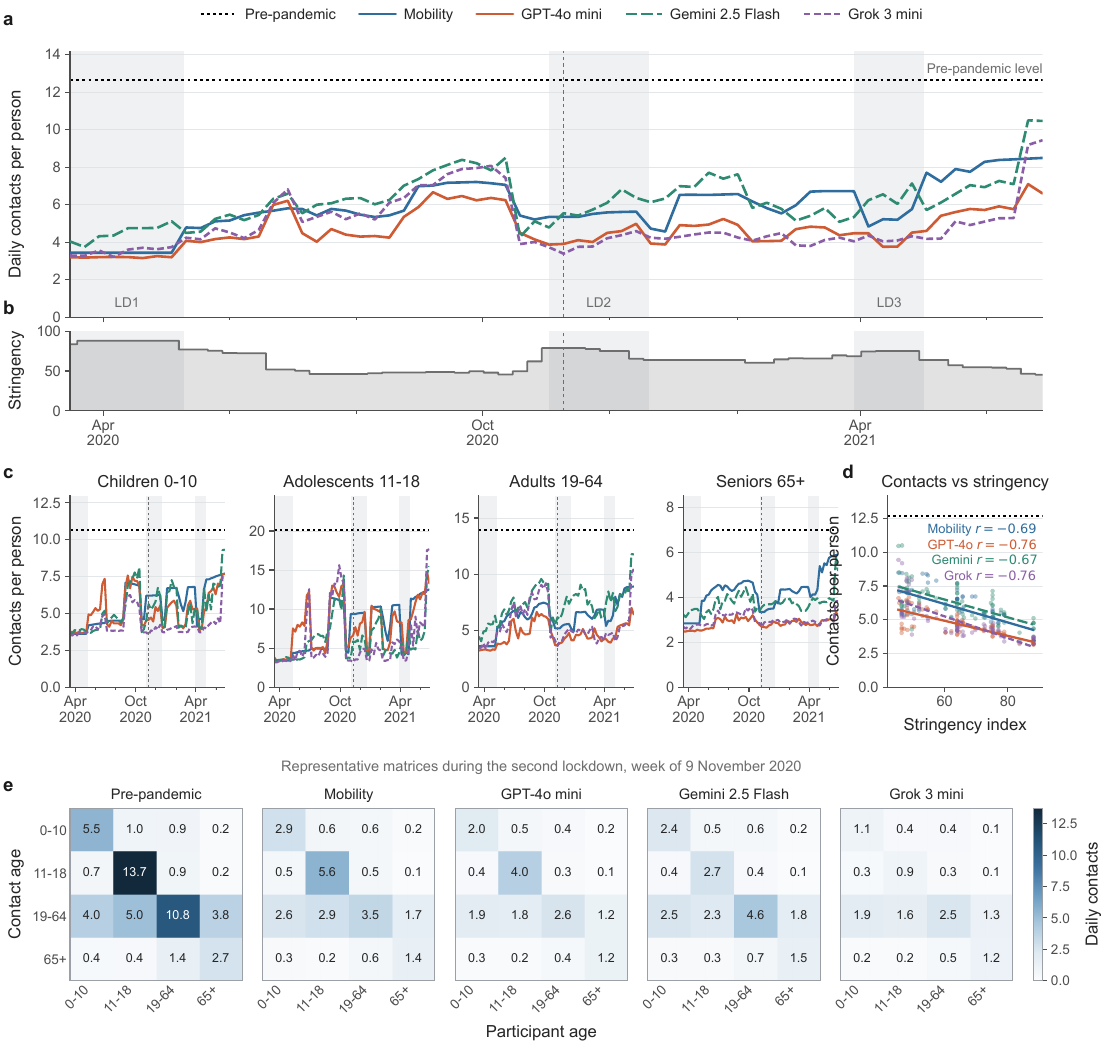}
\vspace{-5mm}
\caption{\textbf{\model generates adaptive and age-specific contact patterns.}
(\textbf{a}) Population-weighted mean number of daily contacts per person from
March 2020 to July 2021 for the three LLMs and the
mobility-driven matrices. The horizontal dotted line denotes the
pre-pandemic baseline. Shaded intervals denote the three national lockdowns,
and the vertical dashed line identifies the week beginning 9 November 2020.
(\textbf{b}) Weekly Oxford COVID-19 government policy stringency
index.
(\textbf{c}) Mean daily contacts per person among children aged 0--10 years,
adolescents aged 11--18 years, adults aged 19--64 years, and seniors aged
65 years or older. Horizontal dotted lines denote the corresponding
pre-pandemic values. Vertical axis limits differ among age groups.
(\textbf{d}) Weekly population contact intensity plotted against the stringency
index. Lines show ordinary least-squares fits, and annotations report Pearson
correlation coefficients.
(\textbf{e}) Reciprocity-corrected contact matrices for all five matrix configurations in the week beginning 9 November 2020, during the second national lockdown. Matrix entries are daily contacts per person
and share a common color scale.}
\label{fig:matrices}
\end{figure}

All three tested LLMs consistently translated evolving epidemic and policy conditions into adaptive contact patterns. For each week from March 2020 to July 2021, the three LLMs re-evaluated the contacts recorded in each pre-pandemic diary under the control measures and epidemic conditions. The retained contacts were aggregated into weekly age-structured matrices and compared with the static pre-pandemic matrix (see the ``Contact-matrix construction'' section in the Supplementary Materials) and the mobility-driven synthetic matrices available for the same period (Fig.~\ref{fig:matrices}). 
As shown in Fig.~\ref{fig:matrices}a, from a pre-pandemic level of
$12.67$ contacts per person per day, contact intensity fell during the first
lockdown to $3.15$--$3.74$ across the LLMs, closely bracketing the value of $3.44$ in
the mobility-driven matrices. Contact intensity subsequently recovered as restrictions eased and contracted again during the second national lockdown, when the mean contact intensity ranged from $3.98$ to $5.83$ across the LLMs. The resulting trajectories showed broad temporal agreement with the mobility-driven matrix, strongest for GPT-4o mini ($r=0.85$), followed by Gemini 2.5 Flash ($r=0.74$) and Grok 3 mini ($r=0.60$). Thus, the three LLMs recovered a common sequence of population-level behavioral changes while retaining systematic differences in their overall contact levels.

The generated changes were both heterogeneous across age groups and policy-responsive
(Fig.~\ref{fig:matrices}b--d). During the second lockdown, adolescent contacts fell by
$62.1\%$--$80.1\%$ across the LLMs, compared with $53.0\%$ in
the mobility-driven matrices, whereas adult contacts fell by
$51.8\%$--$70.3\%$, compared with $61.2\%$ in the mobility-driven series. Across
the study period, adolescents experienced the largest maximum reduction in
every LLM configuration, whereas seniors experienced the smallest. At the
population level, contact intensity declined as policy stringency increased,
with correlations of $-0.67$ to $-0.76$ across the LLMs, comparable
to $-0.69$ for the mobility-driven matrices. Similar aggregate contact intensities did not imply equivalent age structure.
In the week beginning 9 November 2020, Gemini 2.5 Flash retained more adult--adult mixing but less
adolescent--adolescent mixing than the mobility-driven matrix, with corresponding
entries of $4.6$ and $2.7$, compared with $3.5$ and $5.6$ in the mobility-driven matrix (Fig.~\ref{fig:matrices}e). GPT-4o mini and Grok 3 mini instead generated lower contact
volumes among adults and seniors while retaining the principal diagonal and
adult-linked blocks.

These differences persisted across all matched weeks (fig.~S2). GPT-4o mini most closely matched the mobility-driven matrices for elements involving children or adolescents but generated lower adult and senior contact intensities. Gemini 2.5 Flash shifted contact intensity toward adults and away from adolescents and seniors, whereas Grok 3 mini generated lower contact intensity across all four age groups. Together, these results show that the behavioral layer recovered common population-level responses to changing epidemic conditions while producing model-specific patterns of age allocation and assortative mixing; we next examined whether this generated structure remained informative when propagated through the epidemiological model.

\subsection*{LLM-generated behavioral structure informs age-specific epidemic dynamics}

We next tested whether the behavioral structure generated by the LLMs remained epidemiologically informative when propagated through the transmission model. Health outcomes were calibrated to daily SI-VIC hospital admissions summed across age groups, while age-stratified admissions were withheld from fitting and reserved for evaluation. Calibration included a time-varying correcting factor that uniformly rescaled each contact matrix while preserving its relative age structure. This design allowed overall transmission intensity to be adjusted without directly fitting how transmission was distributed across age groups, providing a held-out test of the age-specific mixing structure supplied by the behavioral layer. The complete calibration procedure is reported in the ``Model inference'' section in the Supplementary Materials.

\begin{figure}[!h]
\centering
\includegraphics[
  width=\textwidth,
  height=0.72\textheight,
  keepaspectratio
]{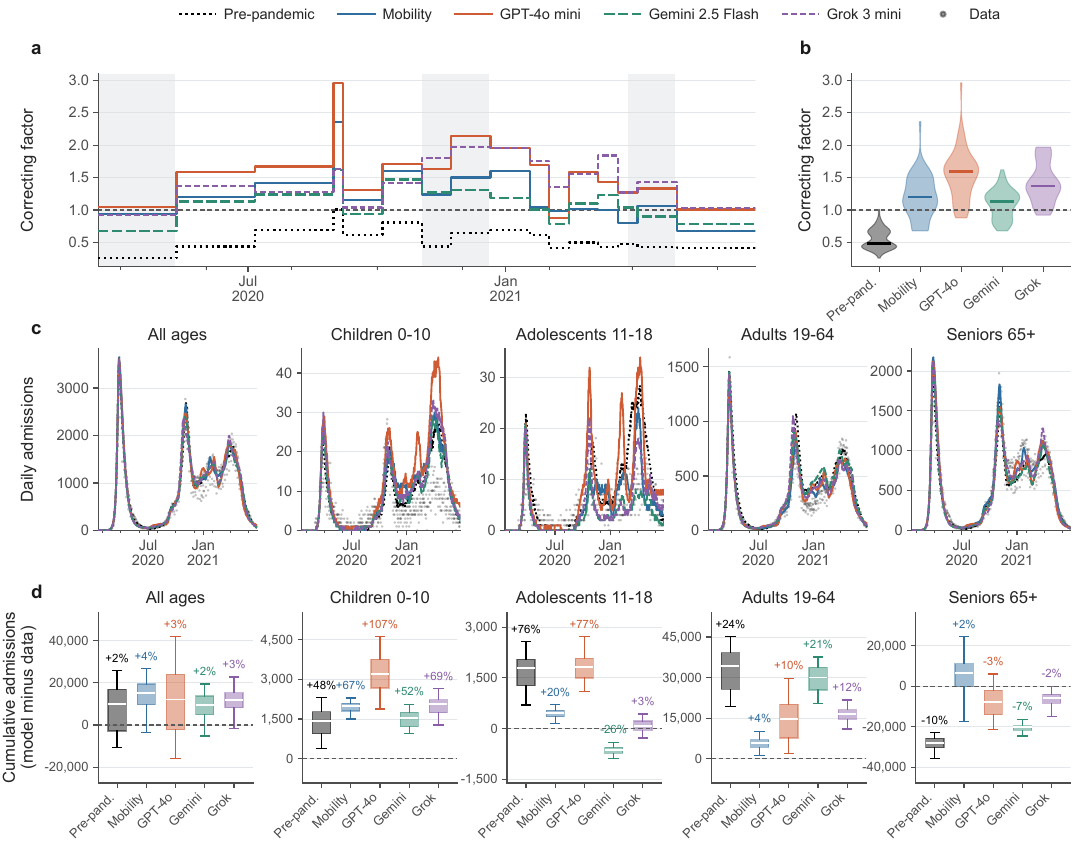}
\caption{\textbf{Retrospective reconstruction of hospital admissions using the
five contact-matrix configurations.}
(\textbf{a}) Time-varying correcting factor, piecewise constant over the
calibration windows. A value of one indicates no additional rescaling of matrix
contact intensity. Shaded intervals denote the three national lockdowns.
(\textbf{b}) Distribution of the correcting factor across fitted weeks. The
factor is piecewise constant within each calibration window. Horizontal bars
denote medians, and the dashed line marks a correcting factor of one.
(\textbf{c}) Observed and simulated daily hospital admissions overall and by
age group. Lines show ensemble medians over 100 stochastic realizations, and points show SI-VIC observations.
(\textbf{d}) Difference between simulated and observed cumulative admissions
over the full study period, overall and by age group. Boxes show distributions
across stochastic realizations, center lines denote medians, boxes span the
interquartile range, and whiskers extend to 1.5 times the interquartile range.
The horizontal dashed line denotes zero difference. Percentage labels report
the median cumulative bias relative to the observed total for the corresponding
age group.}
\label{fig:retrospective}
\end{figure}

The amount of rescaling differed among the configurations
(Fig.~\ref{fig:retrospective}a,b). Gemini 2.5 Flash required the least adjustment among the
time-varying configurations, with a median correcting factor of $1.13$, followed by the mobility-driven matrices at $1.20$, Grok 3 mini at $1.37$, and GPT-4o mini at $1.59$. The static pre-pandemic matrix instead required substantial downward correction, with a median correcting factor of $0.48$. A correcting factor near one indicates that relatively little additional scaling was required to reconcile the contact representation with aggregate hospitalization dynamics. 

After calibration, all five configurations reproduced the timing and aggregate scale of the national hospitalization waves (Fig.~\ref{fig:retrospective}c). Their median cumulative totals were only $2.3\%$--$3.6\%$ above the observed total (Fig.~\ref{fig:retrospective}d). Notably, the static pre-pandemic matrix achieved only a $+2\%$ cumulative bias with substantial downward rescaling. However, it overestimated adult admissions by $24.0\%$ and underestimated senior admissions by $10.2\%$. This illustrates how good aggregate fit can mask misspecified age-specific mixing. The more informative test was therefore their performance on the withheld age-specific admissions. No contact representation consistently outperformed the others across all age groups. The mobility-driven matrices provided a strong retrospective benchmark, particularly for adult and senior admissions, while performance among the LLM-generated matrices varied by model and age group. Nevertheless, GPT-4o mini and Grok 3 mini substantially reduced the adult–senior imbalance produced by the static matrix, with biases of $+10.2\%$ and $-2.9\%$ for GPT-4o mini and $+11.6\%$ and $-2.1\%$ for Grok 3 mini, respectively. Because age-specific admissions were not used for calibration and the correcting factor preserved relative age mixing, these results indicate that, although the LLM-generated matrices did not consistently outperform the mobility-driven benchmark retrospectively, their relative age-specific structure remained epidemiologically informative after aggregate transmission intensity was calibrated.


\subsection*{LLM-based behavioral adaptation improves prospective epidemic forecasts}

\begin{figure}[!h]
\centering
\includegraphics[
  width=\textwidth,
  height=0.72\textheight,
  keepaspectratio
]{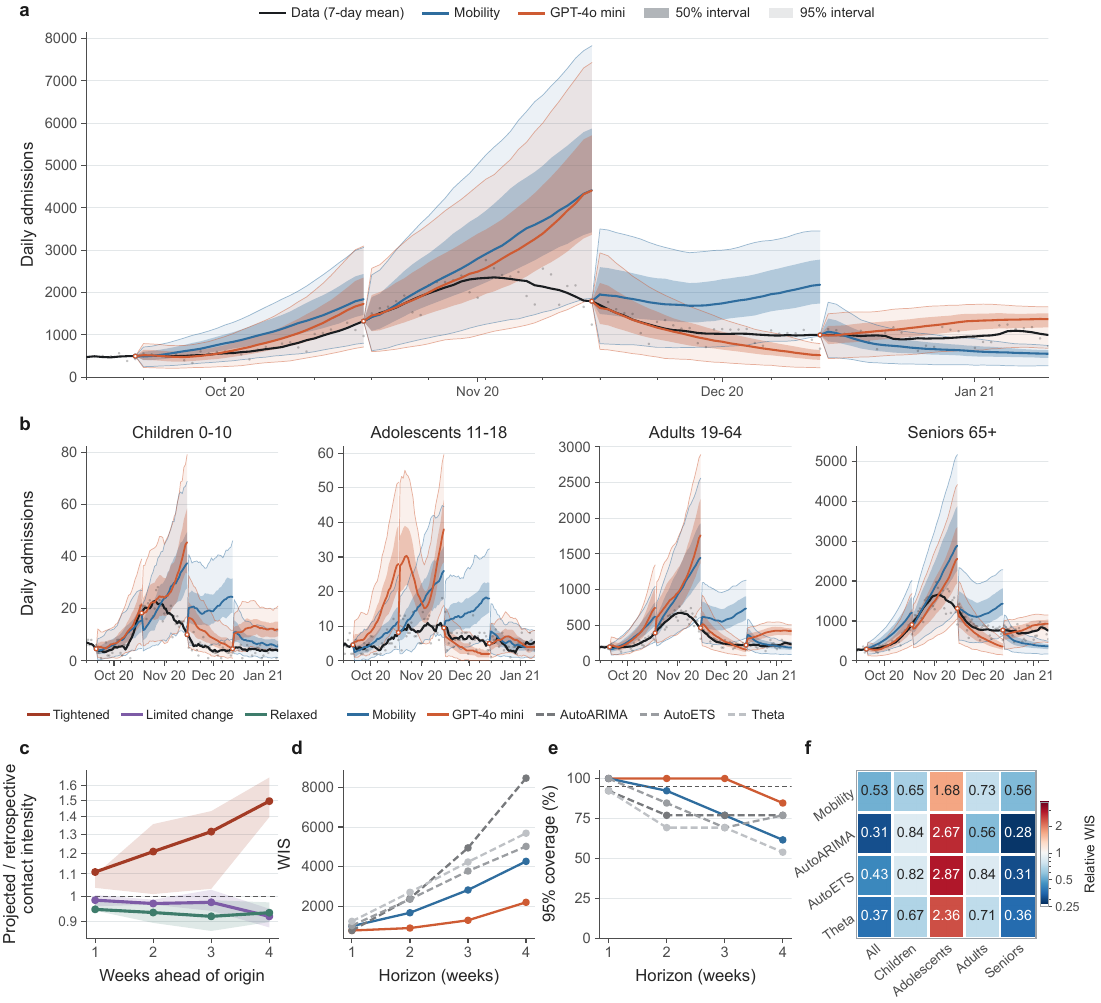}
\caption{\textbf{Real-time forecasts from rolling origins during the second
epidemic wave.}
(\textbf{a}) Observed daily hospital admissions and selected four-week projections from the \model and mobility-driven matrix configurations, summed over age groups. Points show daily observations and the black line shows their seven-day moving average. Lines show predictive medians; darker and lighter
bands show the 50\% and 95\% prediction intervals, respectively. Open symbols mark forecast origins.
(\textbf{b}) Corresponding projections by age group. Shading in
\textbf{a} and \textbf{b} denotes the second national lockdown.
(\textbf{c}) Ratio of projected to retrospectively generated contact intensity by forecast horizon. Origins are classified by the
four-week change in stringency index as tightening, limited change, or relaxation. Lines show means and bands show interquartile ranges.
(\textbf{d}) Mean weighted interval score by forecast horizon.
(\textbf{e}) Mean 95\% prediction-interval coverage; the dashed line denotes the nominal level.
(\textbf{f}) \model's WIS relative to each reference method, overall and by age group. Values below one favor \model.}
\label{fig:forecast}
\end{figure}

We conducted four-week rolling-origin forecasts from 13 consecutive weekly origins, from the week beginning 14 September through the week beginning 7 December 2020, spanning the rise, second national lockdown, and decline of the autumn 2020 wave. At each forecast origin, only information available up to that week was used. LLMs estimated the contact matrix for the next week, the epidemiological model projected the resulting epidemic state, and this projected state was then fed back to the behavioral layer to generate contacts for the following week; this process was repeated iteratively over the four-week horizon (see the ``Real-time forecasting'' section in the Supplementary Materials). For comparison, the mobility-driven configuration used the same epidemiological model but held the most recently observed contact matrix fixed throughout the forecast horizon because future mobility was unavailable. Example forecasts are shown overall and by age group in Fig.~\ref{fig:forecast}a,b.

LLM-generated contact intensity remained close to the corresponding retrospective reconstruction (Fig.~\ref{fig:matrices}a) when policy conditions remained stable after the forecast origin, but diverged when policies subsequently changed (Fig.~\ref{fig:forecast}c).
We classified forecast origins according to the four-week change in policy stringency index: tightened (increase of at least 10 points), relaxed (decrease of at least 10 points), or limited change (absolute change below 10 points). Under the limited change scenario, projected contact intensity differed from the retrospective estimate by less than 3.0\% through two weeks. By contrast, under the tightened scenario, projected contacts increasingly exceeded the retrospective values, with the mean ratio rising from 1.11 at one week to 1.50 at four weeks. Thus, the largest discrepancies in prospective behavioral prediction arose when future policy changes were not yet known.

The LLM-based behavioral layer improved short-term hospitalization forecasts, with the largest gains emerging at longer horizons. \model achieved the lowest mean weighted interval score (WIS) overall, at $1298$, compared with $2443$ for the mobility-driven configuration and $3047$--$4148$ for the three statistical references (Fig.~\ref{fig:forecast}d,f). At one-week out forecast, its performance was comparable to the strongest statistical benchmark (WIS $782$ versus $781$ for automatic ARIMA), whereas from two to four weeks \model consistently outperformed every reference model, with a relative WIS of $0.46$--$0.54$ against the mobility-driven configuration, corresponding to a $46\%$--$54\%$ reduction in WIS. \model also achieved $96.2\%$ coverage of the $95\%$ prediction intervals, compared with $71.2\%$--$82.7\%$ across the reference methods (Fig.~\ref{fig:forecast}e), and produced the lowest WIS for three of the four age groups evaluated (Fig.~\ref{fig:forecast}f).

The largest \model errors occurred at weeks beginning 12 October and 26 October 2020, whose horizons
included the introduction of the second national lockdown. These weeks also had the largest projected-to-retrospective contact ratios, $1.51$ and $1.38$,
and substantial positive admission errors. Across origins, \model
outperformed the mobility-driven configuration at 11 of 13 origins. The remaining failures were
therefore concentrated in forecasts that carried pre-lockdown policy conditions
forward after restrictions had tightened.


\subsection*{\model enables prospective evaluation of alternative policy strategies}

\begin{figure}[p]
\centering
\includegraphics[
  width=\textwidth,
  height=0.72\textheight,
  keepaspectratio
]{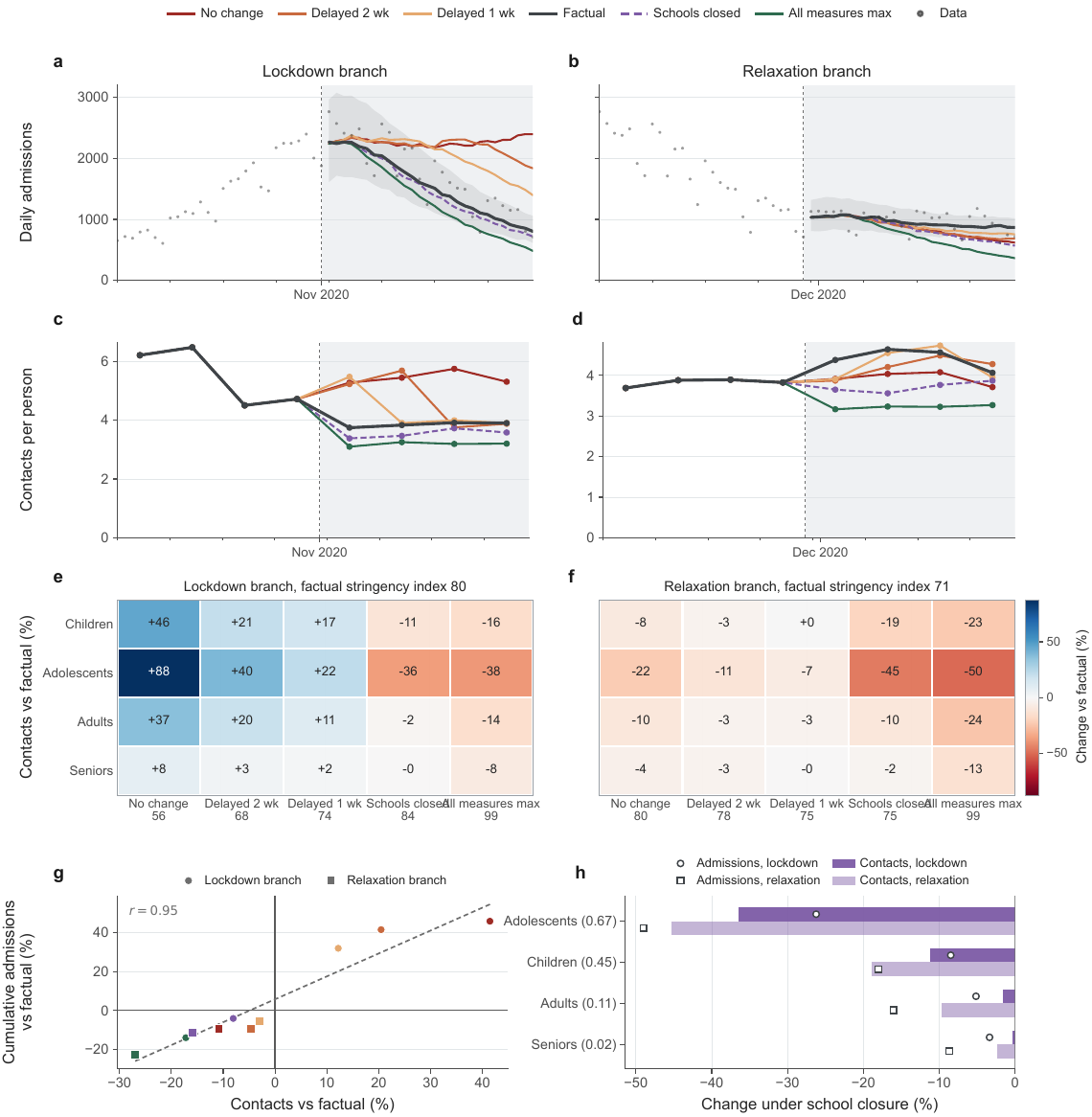}
\caption{\textbf{Policy-conditioned and counterfactual projections at the
imposition and relaxation of the second national lockdown.}
Vertical dashed lines mark the branch points, and shading marks the four-week projection windows.
(\textbf{a}, \textbf{b}) Observed daily hospital admissions and ensemble-median projections under six policy scenarios. The gray band is the interquartile range of the factual-policy projection.
(\textbf{c}, \textbf{d}) Aggregate contact intensity under the corresponding scenarios.
(\textbf{e}, \textbf{f}) Mean four-week age-specific contact change relative to the factual scenario. Numbers below scenario labels are mean Oxford stringency indices; panel titles give the factual values.
(\textbf{g}) Four-week cumulative admission change versus aggregate contact change, relative to the factual scenario. Symbols distinguish branch points, colors distinguish scenarios, and the dashed line is a least-squares fit.
(\textbf{h}) Age-specific changes under school closure. Bars show contact intensity and open symbols show cumulative admissions; parentheses give the pre-pandemic share of contacts occurring at school.}
\label{fig:counterfactual}
\end{figure}

We next evaluated whether \model could support prospective policy analysis by simulating how alternative interventions reshape behavior and epidemic outcomes. At each projected week, the candidate policy and evolving epidemic state informed the LLM behavioral layer, which generated an age-structured contact matrix that drove the next epidemic state; this behavior--epidemic feedback was iterated over the projection horizon. We considered two contrasting decision points: 1) immediately before the second national lockdown and 2) immediately before the subsequent relaxation. At each point, the implemented policy scenario was compared with alternatives that varied intervention timing (one- and two-week policy delays) and intensity (school closure and maximally stringent measures). Epidemic history, fitted parameters, correcting factors, and stochastic seeds were held fixed across scenarios, so differences in projected outcomes arose from the alternative policy scenarios (Fig.~\ref{fig:counterfactual}). Detailed scenario definitions and implementation are provided in the ``Counterfactual policy experiments'' section in the Supplementary Materials, and in tables~S4 and~S5.
This design tests whether \model can prospectively compare the behavioral and epidemiological consequences of candidate policies before implementation.

Because counterfactual outcomes cannot be observed, we first evaluated the framework under the factual policy scenario. This provided an empirical test of whether \model could generate prospective epidemic trajectories when the future policy sequence was specified. Across 13 rolling origins, mean absolute relative error increased from $2.8\%$ at one week to $21.7\%$ at four weeks, with $96\%$ coverage of the 95\% prediction intervals. At the two policy decision points subsequently used for scenario analysis, cumulative admission biases over the four-week projection windows were $-9.7\%$ and $-2.9\%$, respectively (fig.~S3; see the ``Validation under the factual policy scenario'' section in the Supplementary Materials). These factual-policy projections thus provided an empirically evaluated reference for comparison with the alternative policy scenarios.

Alternative policy timing produced coherent and ordered changes in both behavior and epidemic burden. At the lockdown branch, retaining the measures in force in late October produced $41.4\%$ more contacts and $45.8\%$ more cumulative admissions over the following four weeks than the factual scenarios. Delaying the lockdown by one or two weeks also increased both outcomes, with larger effects for the longer delay. Conversely, at the relaxation branch, maintaining the pre-relaxation measures reduced contacts by $10.8\%$ and admissions by $9.6\%$, and delaying relaxation produced smaller reductions in the same direction (Fig.~\ref{fig:counterfactual}a--f). These results show that \model translated the timing of candidate interventions into corresponding behavioral and epidemic consequences.

\model also distinguished policy intensity and composition. Maximum measures reduced contacts and admissions by $17.2\%$ and $14.0\%$ at the lockdown branch and by $27.0\%$ and $22.7\%$ at the relaxation branch. School closure alone produced
smaller aggregate effects. At the relaxation branch, however, school closure and a one-week
delay in relaxation had the same mean stringency index of $74.5$ but generated
substantially different responses: contact intensity declined by $15.9\%$ and
$2.9\%$, and cumulative admissions by $11.4\%$ and $5.3\%$, respectively. These differences were also age-specific (Fig.~\ref{fig:counterfactual}e,f,h), where school closure reduced adolescent contacts by $45.3\%$ but senior contacts by only $2.4\%$, reflecting the different activities targeted by the intervention. 

Across alternative scenarios, changes in behavior translated consistently into changes in epidemic outcomes. Across the ten non-factual scenarios, changes in aggregate contact intensity
were strongly associated with changes in cumulative hospitalization admissions
($r=0.95$; Fig.~\ref{fig:counterfactual}g). By integrating policy content, epidemic conditions, and age-specific activity context, the LLM behavioral layer generated distinct responses to policy scenarios that had not been observed and propagated them through the evolving epidemic. This capability distinguishes \model from approaches that rely on realized behavioral observations or prespecified response functions, enabling prospective model-based evaluation of candidate interventions before implementation.

\section*{Discussion}


\model provides a flexible mechanism for generating adaptive behavioral responses within an evolving epidemic system. It does not require contemporaneous behavioral measurements, and unlike predefined behavioral models, it does not require separate response functions for each contextual factor or interaction. By using LLMs as a context-sensitive behavioral layer, the framework translates changing epidemic and policy conditions into transmission-relevant responses and updates those responses as conditions evolve. The resulting behavioral representations are coupled directly to the epidemiological model, creating a dynamic feedback process in which epidemic conditions shape behavior, behavior modifies transmission, and the resulting epidemic trajectory informs subsequent behavioral adaptation.


The challenge in modeling adaptive behavior lies not in any single behavioral driver, but in how multiple contextual influences interact to shape human response. LLMs provide a flexible way to represent these interactions by jointly interpreting heterogeneous information within a common behavioral context, allowing \model to generate responses that are sensitive to specific conditions and new combinations of those conditions. Crucially, these responses are translated into structured, transmission-relevant representations rather than used as unconstrained free-form outputs. In our implementation, age-structured contact matrices provide this interface between contextual behavioral generation and mechanistic transmission. The three LLMs produced similar broad temporal patterns of behavioral adaptation but differed systematically in age-specific contact structure, reflecting model-specific differences in how contextual cues were translated into behavioral responses. These structural differences propagated into age-specific hospitalization dynamics even after aggregate transmission intensity was calibrated, showing that behavioral adaptation affects not only how much populations interact, but also who interacts with whom. The resulting contact structure therefore carries epidemiologically meaningful information beyond overall contact intensity.

Generative behavioral modeling becomes most valuable prospectively, when future behavioral responses have not yet been observed. Mobility and contact surveys provide strong retrospective signals, but cannot directly specify how populations will respond to future conditions or interventions. \model addresses this gap by generating behavioral responses from the information available at the time and updating them as epidemic conditions evolve. This capability creates a new interface for decision support, allowing candidate interventions to be evaluated through the behavioral responses they are expected to induce. When supplied with the policies subsequently implemented in France while future hospitalization outcomes were withheld, \model reproduced subsequent epidemic trajectories with useful prospective accuracy. This factual-policy evaluation provides an empirical reference for counterfactual analysis, although it cannot directly validate the effects of unobserved alternative policies. More broadly, the framework could allow decision makers to compare when an intervention is introduced, what behaviors it targets, and which population groups are most affected, while propagating these differences through the epidemic system. Although demonstrated here for COVID-19 and contact behavior, the approach is not conceptually restricted to a particular pathogen, policy, or behavioral representation. Future implementations could extend the behavioral layer to other pathogens and transmission-relevant responses, such as testing, vaccination, or healthcare seeking. In this sense, generative behavioral modeling offers a general strategy for incorporating adaptive human response into prospective epidemic analysis and public-health decision-making.

This study has several limitations. First, the LLM-generated contact matrices could not be directly validated against longitudinal measurements of realized contact behavior over the same period. We therefore evaluated them indirectly against contact patterns reconstructed from observed mobility data (fig.~S2) and against observed hospitalization dynamics after propagation through the epidemiological model. Second, differences among LLMs also indicate uncertainty in how epidemic context is translated into behavior and suggest that models may encode distinct behavioral priors or response tendencies~\cite{xie2026evaluating}. A better understanding of these model-specific preferences will be important when selecting and implementing LLM-based behavioral layers, alongside future ensemble approaches and explicit propagation of behavioral-model uncertainty. 
Third, because the retrospective evaluation uses historical COVID-19
conditions, information acquired during LLM pretraining cannot be excluded
completely. We reduced historical identifiability in the main analysis by
masking calendar dates and the pathogen name, and further examined this concern in a dedicated sensitivity analysis reported in the ``Sensitivity to historical information leakage'' section in the Supplementary Materials. In that analysis, we removed additional identifying cues or, conversely,
explicitly disclosed the historical episode while holding the behavioral task
and downstream modeling pipeline fixed. Generated contact structure and
epidemiological performance remained similar across these conditions (fig.~S4),
suggesting that the main findings were robust to the historical information retained in the baseline prompt. However, these tests cannot exclude an influence of pretraining information on model behavior, and evaluation in genuinely novel epidemic settings remains necessary to establish generalization
beyond historical events that may have been represented during pretraining.
Fourth, counterfactual outcomes remain
conditional model projections because the alternative policy scenarios were not observed. Validation under the factual policy scenarios supports the capability of \model but does not establish the accuracy of any individual counterfactual effect. Finally, the empirical evaluation is limited to COVID-19 in France, and the generalizability of the behavioral layer across pathogens, populations, and policy environments remains to be established.


\clearpage
\section*{Materials and Methods}

\subsection*{Data sources}

The contact matrices were anchored to pre-pandemic contact diaries collected in
a large population-based survey in France \cite{beraud2015french}, accessed
through the SOCRATES data tool \cite{willem2020socrates}. Each diary records a
respondent's age together with the age and setting of every person contacted on
a regular weekday. Contact settings comprise home, work, school, transport,
leisure, and other. We denote the resulting collection of baseline contact
diaries by $\mathcal{D}$.

Government control measures were obtained from the Oxford COVID-19 Government
Response Tracker \cite{hale2021oxcgrt}. For retrospective contact
reconstruction, weekly epidemic information supplied to the behavioral layer
was constructed from national reported cases and deaths
\cite{owid_covid}. In prospective analyses, the epidemic input was constructed
from hospital admissions available before the target week and subsequently
updated using model-generated admissions. 

Daily age-stratified hospital admissions were obtained from the SI-VIC database
maintained by Sant\'e publique France \cite{spf_sivic}. Admissions summed over
age groups were used for calibration, whereas age-stratified admissions were
withheld from fitting and used for evaluation. Population structure was taken
from national age-specific population estimates \cite{insee_population}, and
vaccination inputs were based on administered-dose records
\cite{spf_vacsi}. Individuals were grouped as children aged 0 to 10 years,
adolescents aged 11 to 18 years, adults aged 19 to 64 years, and seniors aged
65 years or older.

\subsection*{GABLE framework}

\model couples a generative behavioral layer to an age-structured stochastic
epidemiological model through a time-varying contact matrix. Changing epidemic
and policy conditions modify population contact structure, and the resulting
contact patterns alter subsequent disease transmission.

Throughout the framework equations, $t$ indexes model time. In the empirical
implementation, behavioral contexts and contact matrices were updated weekly,
while epidemiological transitions were simulated daily with the corresponding
weekly contact matrix held fixed between behavioral updates. Let
$\mathbf{u}(t)$ denote the public-health policy conditions and their duration,
and $\mathbf{e}(t)$ the epidemic information supplied to the behavioral layer.
We write
$\mathbf{h}(t)
=
\left(
\mathbf{u}(t),
\mathbf{e}(t)
\right)$
for the epidemic and policy context under which contact behavior is generated.

The behavioral layer operates on the complete baseline contact diary of each
participant. For participant $p$, let $a_p$ denote demographic information and
$\mathbf{x}_p=(x_{p1},\ldots,x_{pK_p})$ the $K_p$ contacts recorded in the
baseline diary. LLM $m$ evaluates the diary under context
$\mathbf{h}(t)$ and returns a vector of contact-retention decisions,
\begin{equation}
\mathbf{d}_p(t)
\sim
\mathcal{B}_m
\left(
\cdot
\mid
a_p,
\mathbf{x}_p,
\mathbf{h}(t)
\right),
\qquad
\mathbf{d}_p(t)\in\{0,1\}^{K_p},
\label{eq:contact_decision}
\end{equation}
where component $d_{pk}(t)=1$ indicates that baseline contact $k$ is retained.

The binary decisions are subsequently adjusted for deterministic calendar and
household constraints and aggregated using the common survey-reconstruction
procedure,
\begin{equation}
\mathbf{C}(t)
=
\mathcal{M}
\left[
\mathcal{A}_t
\left(
\{\mathbf{d}_p(t)\}_{p};
\mathcal{D}
\right)
\right].
\label{eq:behavioral_layer}
\end{equation}
Here, $\mathcal{A}_t$ applies the deterministic adjustments associated with
the target period, including academic-holiday and household-contact
constraints, and converts the binary decisions into adjusted contact weights.
The operator $\mathcal{M}$ then aggregates those weighted contacts into an
age-structured contact matrix. The same reconstruction procedure is used for
the generated and pre-pandemic matrices; retaining every baseline contact
recovers the pre-pandemic matrix. Complete prompt templates, deterministic adjustments, and matrix-reconstruction equations are provided in the ``LLM prompts and queries'' and``Contact-matrix construction'' sections in the Supplementary Materials.

We instantiated the epidemiological layer using the age-stratified stochastic
SARS-CoV-2 transmission model of
\cite{didomenico2020idf,didomenico2026natcommun}. Let $\mathbf{X}(t)$ denote
the population epidemiological state and
$\boldsymbol{\lambda}(t)=\{\lambda_i^v(t)\}_{i,v}$ the collection of age- and
strain-specific forces of infection. The rate at which a susceptible individual
in age class $i$ acquires strain $v$ is
\begin{equation}
\lambda^{v}_i(t)
=
\beta\,
\alpha(t)\,
\eta^{v}
s^{v}_i
\sum_{c}\sum_{j}
r_{cj}\,
C^{c}_{ij}(t)\,
\frac{I^{v}_{cj}(t)}{N}.
\label{eq:FOI}
\end{equation}
Here, $\beta$ is the per-contact transmission rate, $\alpha(t)$ is a
time-varying correcting factor, $\eta^v$ is the transmission advantage of
strain $v$, $s_i^v$ is age- and strain-specific susceptibility, $r_{cj}$ is
relative infectiousness, and $I^v_{cj}(t)$ is the infectious population in age
class $j$, infectious state $c$, and strain $v$.

The behavioral layer supplies the population contact matrix
$\mathbf{C}(t)=\{C_{ij}(t)\}$. Within the transmission model, this matrix is
mapped to $C^c_{ij}(t)$ by applying the contact reductions associated with
symptoms and case detection for infectious state $c$. Conditional on the
resulting force of infection, the epidemiological state evolves according to
\begin{equation}
\mathbf{X}(t+\Delta t)
=
\Phi
\left(
\mathbf{X}(t),
\boldsymbol{\lambda}(t),
\boldsymbol{\psi};
\boldsymbol{\xi}(t)
\right),
\label{eq:epidemiological_layer}
\end{equation}
where $\Phi$ denotes the stochastic epidemiological process,
$\boldsymbol{\psi}$ contains the epidemiological parameters,
$\boldsymbol{\xi}(t)$ represents stochastic transition variation, and
$\Delta t=1$ day in the numerical implementation. The model is stratified by
age, viral strain, and vaccination status and includes latent,
presymptomatic, asymptomatic, symptomatic, hospital, and recovered states.
Transitions were simulated by $\tau$-leaping. The complete compartment structure is shown in fig.~S1 and the parameter values are given in tables~S2 and~S3 (see the ``Transmission model'' section in the Supplementary Materials).

In prospective simulations, epidemic and behavioral dynamics are coupled
recursively. At successive behavioral updates, the evolving epidemic trajectory
determines the epidemic information supplied to the behavioral layer, which
generates the next contact matrix and thereby modifies subsequent transmission:
\begin{equation}
\mathbf{X}(0:t)
\longrightarrow
\mathbf{e}(t+1)
\longrightarrow
\mathbf{h}(t+1)
\longrightarrow
\mathbf{C}(t+1)
\longrightarrow
\boldsymbol{\lambda}(t+1)
\longrightarrow
\mathbf{X}(t+1).
\label{eq:feedback}
\end{equation}
In retrospective reconstruction, $\mathbf{e}(t)$ contains the surveillance
information observed in the period being reconstructed. In prospective
forecasting and policy analyses, it contains hospital-admission information
available before the target period and is updated recursively using the model's
preceding projections. The behavioral and epidemiological operators remain
unchanged across these settings.

The pathogen was described to the LLMs as a febrile respiratory
infectious disease without identifying COVID-19. We instantiated
$\mathcal{B}_m$ independently with GPT-4o mini, Gemini 2.5 Flash, and Grok 3
mini using the same prompt structure and matrix-construction pipeline. Model settings, complete prompts, response validation, caching, and reproducibility procedures are provided in the ``LLM prompts and queries'' section in the Supplementary Materials, and the fields of the prompt that differ between settings are summarized in table~S1.

\subsection*{Model configurations for comparative evaluation}

To isolate the contribution of the behavioral representation, all
configurations used the same epidemiological structure, natural-history
parameters, vaccination and variant inputs, calibration targets, and simulation
procedure. They differed in the source of the age-structured contact matrix
supplied to Eq.~\ref{eq:FOI}:
\begin{equation}
\mathbf{C}(t)
\in
\left\{
\mathbf{C}^{\mathrm{static}},
\mathbf{C}^{\mathrm{mob}}(t),
\mathbf{C}^{\mathrm{LLM},m}(t)
\right\}.
\label{eq:matrix_arms}
\end{equation}

The static configuration used the pre-pandemic matrix reconstructed from the
French contact survey throughout the study period. The mobility-driven
configuration used the published weekly mobility-driven contact matrices of
\cite{didomenico2026natcommun}, which translate observed population mobility
into time-varying age-structured contact patterns. The three \model
configurations generated $\mathbf{C}^{\mathrm{LLM},m}(t)$ independently with GPT-4o mini, Gemini 2.5 Flash, or Grok 3 mini from the same baseline diaries and weekly contextual information. The five configurations therefore shared the downstream epidemiological model and differed in the representation of population contact.

\subsection*{Model calibration and inference}

Calibration proceeded in two stages. We first estimated the per-contact transmission rate $\beta$ during the pre-lockdown period, when every configuration used the pre-pandemic contact matrix. Candidate values were compared using a simulation-based Poisson objective applied to daily hospital admissions summed over age groups.

With $\beta$ fixed, we estimated a time-varying correcting factor
\cite{didomenico2026natcommun}. The factor applies a common scalar multiplier to all matrix elements, changing the overall contact scale while preserving relative age mixing. It accounts for residual time-varying transmission influences not represented by the contact matrices. The factor was fitted sequentially over predefined multi-week windows spanning the major policy and epidemic phases and was piecewise constant within each window.

For each candidate parameter value, the stochastic model was simulated
repeatedly and the daily ensemble median was used as the Poisson mean. The objective was used for numerical parameter selection and was not interpreted as a full likelihood for the stochastic transmission process. Age-stratified admissions were not used during fitting. The parameter grids, fitting windows, reporting adjustments, and numerical search procedures are provided in the ``Model inference'' section in the Supplementary Materials.

\subsection*{Real-time forecasting}

We evaluated four-week forecasts from 13 rolling origins, the weeks beginning 14 September to 7 December 2020, covering the rise, second national lockdown, and decline of the autumn 2020
epidemic wave. To limit the number of queries, all prospective analyses (real-time forecasting, policy-conditioned validation, and counterfactual projection) used GPT-4o mini as the behavioral layer.


For the first projected week, the epidemic component of the behavioral context
used hospital admissions available through the forecast origin. For later projected weeks, it was updated using the model's own preceding admission projections. Each generated contact matrix entered the transmission model, and
the projected epidemic trajectory informed the subsequent behavioral update,
iterating the feedback relation in Eq.~\ref{eq:feedback} over the four-week
horizon. No hospital admission observed after the forecast origin was used to
construct a future behavioral context.
Future policy changes were treated as unknown in this forecasting experiment.
The seven policy indicators were therefore held at their origin-week levels,
while the policy-duration counter was extended through the forecast horizon.

Correcting factors were fitted weekly through the forecast origin and frozen
forward at the mean of the four most recent fitted values. The mobility-driven configuration used the same transmission and calibration procedure. Its published matrices were used through the forecast origin, and the origin-week matrix was held fixed over the future horizon because subsequent mobility observations were unavailable. Each matrix configuration produced daily age-resolved trajectories from an ensemble of 100 stochastic realizations.

Three statistical forecasting methods provided additional references:
automatic ARIMA \cite{hyndman2008forecast}, automatic exponential smoothing
\cite{hyndman2002ets}, and the Theta method
\cite{assimakopoulos2000theta}. Each was fitted at every origin using complete
weekly hospital admissions available through that origin. Forecast implementation and statistical-baseline specifications are provided in the ``Real-time forecasting'' and ``Statistical forecast baselines'' sections in the Supplementary Materials.

\subsection*{Policy-conditioned validation and counterfactual projections}

The forecasting and policy analyses used the same coupled model and differed in
the future policy scenario supplied to the behavioral layer. For an origin $o$
and projection horizon $H$, let
\begin{equation}
\mathbf{U}^{(s)}_{o+1:o+H}
=
\left\{
\mathbf{u}^{(s)}_{o+1},
\ldots,
\mathbf{u}^{(s)}_{o+H}
\right\}
\label{eq:policy_scenario}
\end{equation}
denote the future policy sequence under scenario $s$.

In real-time forecasting, the policy observed at the origin was carried
forward through the projection horizon because subsequent policy changes were
unknown. In policy-conditioned validation, the policy sequence subsequently
implemented was supplied for each projected week, while future hospital
admissions remained unavailable. Counterfactual projections replaced this
factual scenario with candidate alternatives. In all three prospective settings,
the epidemic component of the behavioral context was updated recursively using
the model's preceding admission projections.

Because epidemic outcomes under alternative policies cannot be observed
directly, we first evaluated the step of the framework that remains empirically
testable: whether GABLE could reproduce subsequent epidemic dynamics when
supplied with the policy sequence that was actually implemented. We applied this
policy-conditioned configuration to the same 13 rolling origins used for
real-time forecasting. At each origin, future policy was supplied, but no
hospital admission was observed after the origin was used. This analysis assessed
whether the coupled behavioral and epidemiological model could generate
prospective trajectories under a known policy scenario before the framework was
used to compare unobserved alternatives.

We then conducted counterfactual experiments at two branch points, immediately
before the second national lockdown and immediately before its subsequent
relaxation. At each branch point, six scenarios were evaluated: the factual
sequence, no policy change after the branch, the factual change delayed by one week, the factual change delayed by two weeks, maximum containment measures, and school closure. All scenarios shared the epidemic history, transmission parameters, vaccination and variant inputs, correcting factor, academic calendar, and stochastic seeds up to the branch. Beginning after the branch, the behavioral layer received the scenario-specific policy sequence, and the resulting contact and epidemic trajectories evolved recursively along that scenario.

The two branch points were selected from the rolling forecast origins using the difference in projected admissions at the four-week target between the
factual-policy and frozen-policy configurations. This identified the week beginning 26 October 2020 as the branch preceding the strongest simulated effect of subsequent policy tightening (the lockdown branch)
and the week beginning 23 November 2020 as the branch preceding the strongest simulated effect of subsequent policy relaxation (the relaxation branch). The complete selection criterion, candidate
origins, policy sequences, policy-duration rules, and Oxford stringency-index calculations are provided in the ``Counterfactual policy experiments'' section in the Supplementary Materials, with candidate origins listed in table~S4 and scenario levels in table~S5.

\subsection*{Evaluation}

Generated contact matrices were first evaluated independently of
hospitalization outcomes.
Aggregate contact intensity was summarized by its minimum during the first national lockdown and its mean during the second national lockdown. Each LLM series was compared with the mobility-driven series over the 68 matched weeks using the Pearson temporal correlation.

Age-specific contact intensities were summarized by their reductions from the
pre-pandemic values during the second lockdown and by the largest reduction
over the study period. Their association with public-health restrictions was
evaluated using the Pearson correlation between weekly aggregate contact
intensity and the Oxford stringency index. Matrix structure was examined in the week beginning 9 November 2020 using reciprocity-corrected matrices displayed on a common scale.
Element-wise comparisons, within-group contact fractions, and age-specific
contact-intensity ratios across all matched weeks are reported in the ``Structural comparison of LLM and mobility-driven contact matrices'' section in the Supplementary Materials, and in fig.~S2.

Retrospective transmission performance was evaluated through the fitted
correcting factor and the resulting hospital admissions. The correcting factor
was summarized by its distribution across fitted weeks and its median, with a
value near one indicating little additional aggregate rescaling. Because
$\alpha(t)$ rescales the complete contact matrix uniformly, differences in
relative age mixing are preserved during calibration.

Hospital admissions were evaluated overall and by age group. Calibration used
admissions summed across age groups, whereas age-stratified admissions were
withheld from fitting. Agreement with these withheld series therefore assessed
the epidemiological consequences of the age-specific mixing structure supplied
by each contact representation after aggregate transmission intensity had been
calibrated. Cumulative bias was calculated by comparing simulated and observed
admissions summed over the evaluation period.

Forecast performance was evaluated over 13 rolling origins and four forecast
horizons, giving 52 matched origin--horizon targets for each method. The target
was total hospital admissions in each complete Monday-to-Sunday week. For the
mechanistic configurations, weekly predictive distributions were obtained by
summing each stochastic daily trajectory over the target week and calculating
quantiles across trajectories.

The primary probabilistic metric was the weighted interval score
\cite{bracher2021wis}, defined in the ``Real-time forecasting'' section
in the Supplementary Materials.
We additionally reported mean absolute error, mean absolute relative error,
cumulative bias, and 95\% prediction-interval coverage. Forecasts were
summarized by horizon, age group, and origin. Relative WIS was calculated as
the mean WIS of GPT-4o mini divided by the mean WIS of the reference method
over matched origin--horizon targets.

To evaluate prospective contact generation, aggregate contact intensity
generated within each forecast horizon was compared with the contact intensity
reconstructed retrospectively for the same calendar week. Origins were
classified according to the change in Oxford stringency index over the
four-week horizon as subsequent tightening, subsequent relaxation, or limited
policy change.

Policy-conditioned performance was summarized using mean absolute relative
error, signed relative error, 95\% interval coverage, and cumulative bias.
This analysis provided the empirical reference for interpreting the subsequent
counterfactual projections. Extended results across all rolling origins are reported in the ``Validation under the factual policy scenario'' section in the Supplementary Materials, and in fig.~S3.

Counterfactual scenarios were compared with the factual-policy projection at
the same branch point. Aggregate and age-specific contact intensities were
averaged over the four projected weeks, and hospital admissions were summed
over the same period. Changes were expressed relative to the factual-policy
projection at the corresponding branch point. We additionally examined the ordering of contact and admission outcomes by mean four-week policy stringency, compared the age-specific responses to school closure with the pre-pandemic school-contact shares, and calculated the Pearson correlation between scenario-induced changes in aggregate contact intensity and cumulative hospital admissions across the ten non-factual scenarios.



\clearpage 

%
\bibliography{science_template} 

@article{didomenico2020idf,
  author  = {Di Domenico, Laura and Pullano, Giulia and Sabbatini, Chiara E. and Bo{\"e}lle, Pierre-Yves and Colizza, Vittoria},
  title   = {Impact of lockdown on {COVID-19} epidemic in {\^I}le-de-France and possible exit strategies},
  journal = {BMC Medicine},
  volume  = {18},
  pages   = {240},
  year    = {2020},
  doi     = {10.1186/s12916-020-01698-4}
}

@article{didomenico2026natcommun,
  author  = {Di Domenico, Laura and Bosetti, Paolo and Sabbatini, Chiara E. and Opatowski, Lulla and Colizza, Vittoria},
  title   = {Mobility-driven synthetic contact matrices as a scalable solution for real-time pandemic response modeling},
  journal = {Nature Communications},
  volume  = {17},
  pages   = {1845},
  year    = {2026},
  doi     = {10.1038/s41467-026-68557-3}
}

@article{bracher2021wis,
  author  = {Bracher, Johannes and Ray, Evan L. and Gneiting, Tilmann and Reich, Nicholas G.},
  title   = {Evaluating epidemic forecasts in an interval format},
  journal = {PLOS Computational Biology},
  volume  = {17},
  number  = {2},
  pages   = {e1008618},
  year    = {2021},
  doi     = {10.1371/journal.pcbi.1008618}
}

@article{davies2020age,
  author  = {Davies, Nicholas G. and Klepac, Petra and Liu, Yang and Prem, Kiesha and Jit, Mark and Eggo, Rosalind M.},
  title   = {Age-dependent effects in the transmission and control of {COVID-19} epidemics},
  journal = {Nature Medicine},
  volume  = {26},
  pages   = {1205--1211},
  year    = {2020},
  doi     = {10.1038/s41591-020-0962-9}
}

@article{beraud2015french,
  author  = {B{\'e}raud, Guillaume and Kazmercziak, Sabine and Beutels, Philippe and Levy-Bruhl, Daniel and Lenne, Xavier and Mielcarek, Nathalie and Yazdanpanah, Yazdan and Bo{\"e}lle, Pierre-Yves and Hens, Niel and Dervaux, Benoit},
  title   = {The {French} connection: the first large population-based contact survey in {France} relevant for the spread of infectious diseases},
  journal = {PLOS ONE},
  volume  = {10},
  number  = {7},
  pages   = {e0133203},
  year    = {2015},
  doi     = {10.1371/journal.pone.0133203}
}

@article{willem2020socrates,
  author  = {Willem, Lander and Van Hoang, Thang and Funk, Sebastian and Coletti, Pietro and Beutels, Philippe and Hens, Niel},
  title   = {{SOCRATES}: an online tool leveraging a social contact data sharing initiative to assess mitigation strategies for {COVID-19}},
  journal = {BMC Research Notes},
  volume  = {13},
  pages   = {293},
  year    = {2020},
  doi     = {10.1186/s13104-020-05136-9}
}

@article{hale2021oxcgrt,
  author  = {Hale, Thomas and Angrist, Noam and Goldszmidt, Rafael and Kira, Beatriz and Petherick, Anna and Phillips, Toby and Webster, Samuel and Cameron-Blake, Emily and Hallas, Laura and Majumdar, Saptarshi and Tatlow, Helen},
  title   = {A global panel database of pandemic policies ({Oxford COVID-19} Government Response Tracker)},
  journal = {Nature Human Behaviour},
  volume  = {5},
  pages   = {529--538},
  year    = {2021},
  doi     = {10.1038/s41562-021-01079-8}
}

@misc{spf_sivic,
  author       = {{Sant{\'e} publique France}},
  title        = {Donn{\'e}es hospitali{\`e}res relatives {\`a} l'{\'e}pid{\'e}mie de {COVID-19} ({SI-VIC})},
  howpublished = {\url{https://www.data.gouv.fr/fr/datasets/donnees-hospitalieres-relatives-a-lepidemie-de-covid-19/}},
  year         = {2020}
}

@misc{insee_population,
  author       = {{Institut national de la statistique et des {\'e}tudes {\'e}conomiques (INSEE)}},
  title        = {Estimations de population par r{\'e}gion et d{\'e}partement},
  howpublished = {\url{https://www.insee.fr/fr/statistiques/3696315}},
  year         = {2020}
}

@misc{insee_households,
  author       = {{Institut national de la statistique et des {\'e}tudes {\'e}conomiques (INSEE)}},
  title        = {M{\'e}nages selon leur taille},
  howpublished = {\url{https://www.insee.fr}},
  year         = {2020}
}

@misc{spf_vacsi,
  author       = {{Sant{\'e} publique France}},
  title        = {Donn{\'e}es relatives aux personnes vaccin{\'e}es contre la {COVID-19} ({VAC-SI})},
  howpublished = {\url{https://www.data.gouv.fr/fr/datasets/donnees-relatives-aux-personnes-vaccinees-contre-la-covid-19-1/}},
  year         = {2021}
}

@article{owid_covid,
  author  = {Mathieu, Edouard and Ritchie, Hannah and Rod{\'e}s-Guirao, Lucas
             and Appel, Cameron and Gavrilov, Daniel and Giattino, Charlie
             and Hasell, Joe and Macdonald, Bobbie and Dattani, Saloni
             and Beltekian, Diana and Ortiz-Ospina, Esteban and Roser, Max},
  title   = {{COVID-19} Pandemic},
  journal = {Our World in Data},
  year    = {2020},
  note    = {\url{https://ourworldindata.org/coronavirus}}
}

@article{gaymard2021alpha,
    author  = {Gaymard, Alexandre and Bosetti, Paolo and Feri, Adeline
             and Destras, Gregory and Enouf, Vincent and Andronico, Alessio
             and Burrel, Sonia and Behillil, Sylvie and Sauvage, Claire
             and Bal, Antonin and Morfin, Florence and Van Der Werf, Sylvie
             and Josset, Laurence and {ANRS MIE AC43 COVID-19}
             and {French viro COVID group} and Blanquart, Fran{\c c}ois
             and Coignard, Bruno and Cauchemez, Simon and Lina, Bruno},
  title   = {Early assessment of diffusion and possible expansion of {SARS-CoV-2} Lineage 20I/501Y.V1 ({B.1.1.7}, variant of concern 202012/01) in {France}, January to March 2021},
  journal = {Eurosurveillance},
  volume  = {26},
  number  = {9},
  pages   = {2100133},
  year    = {2021},
  doi     = {10.2807/1560-7917.ES.2021.26.9.2100133}
}

@article{paredes2022hospitalization,
  author  = {Paredes, Miguel I. and Lunn, Stephanie M. and Famulare, Michael
             and Frisbie, Lauren A. and Painter, Ian and Burstein, Roy
             and Roychoudhury, Pavitra and Xie, Hong and Mohamed Bakhash, Shah A.
             and Perez, Ricardo and Lukes, Maria and Ellis, Sean and Sathees, Saraswathi
             and Mathias, Patrick C. and Greninger, Alexander and Starita, Lea M.
             and Frazar, Chris D. and Ryke, Erica and Zhong, Weizhi and Gamboa, Luis
             and Threlkeld, Machiko and Lee, Jover and McDermot, Evan and Truong, Melissa
             and Nickerson, Deborah A. and Bates, Daniel L. and Hartman, Matthew E.
             and Haugen, Eric and Nguyen, Truong N. and Richards, Joshua D.
             and Rodriguez, Jacob L. and Stamatoyannopoulos, John A. and Thorland, Eric
             and Melly, Geoff and Dykema, Philip E. and MacKellar, Drew C.
             and Gray, Hannah K. and Singh, Avi and Peterson, JohnAric M.
             and Russell, Denny and Torres, Laura Marcela and Lindquist, Scott
             and Bedford, Trevor and Allen, Krisandra J. and Oltean, Hanna N.},
  title   = {Associations Between Severe Acute Respiratory Syndrome Coronavirus 2
             ({SARS-CoV-2}) Variants and Risk of Coronavirus Disease 2019 ({COVID-19})
             Hospitalization Among Confirmed Cases in {Washington State}:
             A Retrospective Cohort Study},
  journal = {Clinical Infectious Diseases},
  volume  = {75},
  number  = {1},
  pages   = {e536--e544},
  year    = {2022},
  doi     = {10.1093/cid/ciac279}
}

@article{lauer2020incubation,
  author  = {Lauer, Stephen A. and Grantz, Kyra H. and Bi, Qifang and Jones, Forrest K. and Zheng, Qulu and Meredith, Hannah R. and Azman, Andrew S. and Reich, Nicholas G. and Lessler, Justin},
  title   = {The incubation period of coronavirus disease 2019 ({COVID-19}) from publicly reported confirmed cases: estimation and application},
  journal = {Annals of Internal Medicine},
  volume  = {172},
  number  = {9},
  pages   = {577--582},
  year    = {2020},
  doi     = {10.7326/M20-0504}
}

@article{ferretti2020quantifying,
  author  = {Ferretti, Luca and Wymant, Chris and Kendall, Michelle and Zhao, Lele and Nurtay, Anel and Abeler-D{\"o}rner, Lucie and Parker, Michael and Bonsall, David and Fraser, Christophe},
  title   = {Quantifying {SARS-CoV-2} transmission suggests epidemic control with digital contact tracing},
  journal = {Science},
  volume  = {368},
  number  = {6491},
  pages   = {eabb6936},
  year    = {2020},
  doi     = {10.1126/science.abb6936}
}

@article{lavezzo2020suppression,
  author  = {Lavezzo, Enrico and Franchin, Elisa and Ciavarella, Constanze
             and Cuomo-Dannenburg, Gina and Barzon, Luisa and Del Vecchio, Claudia
             and Rossi, Lucia and Manganelli, Riccardo and Loregian, Arianna
             and Navarin, Nicol{\`o} and Abate, Davide and Sciro, Manuela
             and Merigliano, Stefano and De Canale, Ettore and Vanuzzo, Maria Cristina
             and Besutti, Valeria and Saluzzo, Francesca and Onelia, Francesco
             and Pacenti, Monia and Parisi, Saverio G. and Carretta, Giovanni
             and Donato, Daniele and Flor, Luciano and Cocchio, Silvia and Masi, Giulia
             and Sperduti, Alessandro and Cattarino, Lorenzo and Salvador, Renato
             and Nicoletti, Michele and Caldart, Federico and Castelli, Gioele
             and Nieddu, Eleonora and Labella, Beatrice and Fava, Ludovico
             and Drigo, Matteo and Gaythorpe, Katy A. M. and Brazzale, Alessandra R.
             and Toppo, Stefano and Trevisan, Marta and Baldo, Vincenzo
             and Donnelly, Christl A. and Ferguson, Neil M. and Dorigatti, Ilaria
             and Crisanti, Andrea and {Imperial College COVID-19 Response Team}},
  title   = {Suppression of a {SARS-CoV-2} outbreak in the {Italian} municipality of {Vo'}},
  journal = {Nature}, volume = {584}, number = {7821},
  pages   = {425--429}, year = {2020}, doi = {10.1038/s41586-020-2488-1}
}

@article{riccardo2020epidemiological,
  author  = {Riccardo, Flavia and Ajelli, Marco and Andrianou, Xanthi D.
             and Bella, Antonino and Del Manso, Martina and Fabiani, Massimo
             and Bellino, Stefania and Boros, Stefano and Urdiales, Alberto Mateo
             and Marziano, Valentina and Rota, Maria Cristina and Filia, Antonietta
             and D'Ancona, Fortunato and Siddu, Andrea and Punzo, Ornella
             and Trentini, Filippo and Guzzetta, Giorgio and Poletti, Piero
             and Stefanelli, Paola and Castrucci, Maria Rita and Ciervo, Alessandra
             and Di Benedetto, Corrado and Tallon, Marco and Piccioli, Andrea
             and Brusaferro, Silvio and Rezza, Giovanni and Merler, Stefano
             and Pezzotti, Patrizio and {COVID-19 working group}},
  title   = {Epidemiological characteristics of {COVID-19} cases and estimates of the reproductive numbers 1 month into the epidemic, {Italy}, 28 {January} to 31 {March} 2020},
  journal = {Eurosurveillance},
  volume  = {25},
  number  = {49},
  pages   = {2000790},
  year    = {2020},
  doi     = {10.2807/1560-7917.ES.2020.25.49.2000790}
}

@article{lapidus2021severe,
  author  = {Lapidus, Nathana{\"e}l and Paireau, Juliette and Levy-Bruhl, Daniel and de Lamballerie, Xavier and Severi, Gianluca and Touvier, Mathilde and Zins, Marie and Cauchemez, Simon and Carrat, Fabrice},
  title   = {Do not neglect {SARS-CoV-2} hospitalization and fatality risks in the middle-aged adult population},
  journal = {Infectious Diseases Now},
  volume  = {51},
  number  = {4},
  pages   = {380--382},
  year    = {2021},
  doi     = {10.1016/j.idnow.2020.12.007}
}

@article{cereda2021early,
  author  = {Cereda, Danilo and Manica, Mattia and Tirani, Marcello and Rovida, Francesca
             and Demicheli, Vittorio and Ajelli, Marco and Poletti, Piero
             and Trentini, Filippo and Guzzetta, Giorgio and Marziano, Valentina
             and Piccarreta, Raffaella and Barone, Antonio and Magoni, Michele
             and Deandrea, Silvia and Diurno, Giulio and Lombardo, Massimo
             and Faccini, Marino and Pan, Angelo and Bruno, Raffaele and Pariani, Elena
             and Grasselli, Giacomo and Piatti, Alessandra and Gramegna, Maria
             and Baldanti, Fausto and Melegaro, Alessia and Merler, Stefano},
  title   = {The early phase of the {COVID-19} epidemic in {Lombardy}, {Italy}},
  journal = {Epidemics}, volume = {37}, pages = {100528}, year = {2021},
  doi     = {10.1016/j.epidem.2021.100528}
}

@article{li2020substantial,
  author  = {Li, Ruiyun and Pei, Sen and Chen, Bin and Song, Yimeng and Zhang, Tao and Yang, Wan and Shaman, Jeffrey},
  title   = {Substantial undocumented infection facilitates the rapid dissemination of novel coronavirus ({SARS-CoV-2})},
  journal = {Science},
  volume  = {368},
  number  = {6490},
  pages   = {489--493},
  year    = {2020},
  doi     = {10.1126/science.abb3221}
}

@article{hu2021infectivity,
  author  = {Hu, Shixiong and Wang, Wei and Wang, Yan and Litvinova, Maria
             and Luo, Kaiwei and Ren, Lingshuang and Sun, Qianlai and Chen, Xinghui
             and Zeng, Ge and Li, Jing and Liang, Lu and Deng, Zhihong and Zheng, Wen
             and Li, Mei and Yang, Hao and Guo, Jinxin and Wang, Kai and Chen, Xinhua
             and Liu, Ziyan and Yan, Han and Shi, Huilin and Chen, Zhiyuan
             and Zhou, Yonghong and Sun, Kaiyuan and Vespignani, Alessandro
             and Viboud, C{\'e}cile and Gao, Lidong and Ajelli, Marco and Yu, Hongjie},
  title   = {Infectivity, susceptibility, and risk factors associated with {SARS-CoV-2}
             transmission under intensive contact tracing in {Hunan}, {China}},
  journal = {Nature Communications}, volume = {12}, number = {1},
  pages   = {1533}, year = {2021}, doi = {10.1038/s41467-021-21710-6}
}

@article{franco2022inferring,
  author  = {Franco, Nicolas and Coletti, Pietro and Willem, Lander and Angeli, Leonardo
             and Lajot, Adrien and Abrams, Steven and Beutels, Philippe
             and Faes, Christel and Hens, Niel},
  title   = {Inferring age-specific differences in susceptibility to and infectiousness
             upon {SARS-CoV-2} infection based on {Belgian} social contact data},
  journal = {PLOS Computational Biology}, volume = {18}, number = {3},
  pages   = {e1009965}, year = {2022}, doi = {10.1371/journal.pcbi.1009965}
}

@article{viner2021susceptibility,
  author  = {Viner, Russell M. and Mytton, Oliver T. and Bonell, Chris
             and Melendez-Torres, G. J. and Ward, Joseph and Hudson, Lee
             and Waddington, Claire and Thomas, James and Russell, Simon
             and van der Klis, Fiona and Koirala, Archana and Ladhani, Shamez
             and Panovska-Griffiths, Jasmina and Davies, Nicholas G.
             and Booy, Robert and Eggo, Rosalind M.},
  title   = {Susceptibility to {SARS-CoV-2} infection among children and adolescents
             compared with adults: a systematic review and meta-analysis},
  journal = {JAMA Pediatrics}, volume = {175}, number = {2},
  pages   = {143--156}, year = {2021}, doi = {10.1001/jamapediatrics.2020.4573}
}

@article{boelle2020trajectories,
  author  = {Bo{\"e}lle, Pierre-Yves and Delory, Tristan and Maynadier, Xavier
             and Janssen, C{\'e}cile and Piarroux, Renaud and Pichenot, Marie
             and Lemaire, Xavier and Baclet, Nicolas and Weyrich, Pierre
             and Melliez, Hugues and Meybeck, Agn{\`e}s and Lanoix, Jean-Philippe
             and Robineau, Olivier},
  title   = {Trajectories of hospitalization in {COVID-19} patients:
             an observational study in {France}},
  journal = {Journal of Clinical Medicine}, volume = {9}, number = {10},
  pages   = {3148}, year = {2020}, doi = {10.3390/jcm9103148}
}

@article{pellis2021challenges,
    author  = {Pellis, Lorenzo and Scarabel, Francesca and Stage, Helena B.
             and Overton, Christopher E. and Chappell, Lauren H. K.
             and Fearon, Elizabeth and Bennett, Emma and Lythgoe, Katrina A.
             and House, Thomas A. and Hall, Ian
             and {University of Manchester COVID-19 Modelling Group}},
  title   = {Challenges in control of {COVID-19}: short doubling time and long delay to effect of interventions},
  journal = {Philosophical Transactions of the Royal Society B},
  volume  = {376},
  number  = {1829},
  pages   = {20200264},
  year    = {2021},
  doi     = {10.1098/rstb.2020.0264}
}

@article{dagan2021bnt162b2,
  author  = {Dagan, Noa and Barda, Noam and Kepten, Eldad and Miron, Oren and Perchik, Shay and Katz, Mark A. and Hern{\'a}n, Miguel A. and Lipsitch, Marc and Reis, Ben and Balicer, Ran D.},
  title   = {{BNT162b2} m{RNA} {Covid-19} vaccine in a nationwide mass vaccination setting},
  journal = {New England Journal of Medicine},
  volume  = {384},
  number  = {15},
  pages   = {1412--1423},
  year    = {2021},
  doi     = {10.1056/NEJMoa2101765}
}

@article{haas2021impact,
  author  = {Haas, Eric J. and Angulo, Frederick J. and McLaughlin, John M.
             and Anis, Emilia and Singer, Shepherd R. and Khan, Farid
             and Brooks, Nati and Smaja, Meir and Mircus, Gabriel and Pan, Kaijie
             and Southern, Jo and Swerdlow, David L. and Jodar, Luis
             and Levy, Yeheskel and Alroy-Preis, Sharon},
  title   = {Impact and effectiveness of m{RNA} {BNT162b2} vaccine against {SARS-CoV-2}
             infections and {COVID-19} cases, hospitalisations, and deaths following a
             nationwide vaccination campaign in {Israel}: an observational study using
             national surveillance data},
  journal = {The Lancet}, volume = {397}, number = {10287},
  pages   = {1819--1829}, year = {2021}, doi = {10.1016/S0140-6736(21)00947-8}
}

@article{sheikh2021delta,
  author  = {Sheikh, Aziz and McMenamin, Jim and Taylor, Bob and Robertson, Chris},
  title   = {{SARS-CoV-2} {Delta} {VOC} in {Scotland}: demographics, risk of hospital admission, and vaccine effectiveness},
  journal = {The Lancet},
  volume  = {397},
  number  = {10293},
  pages   = {2461--2462},
  year    = {2021},
  doi     = {10.1016/S0140-6736(21)01358-1}
}

@article{eyre2022effect,
  author  = {Eyre, David W. and Taylor, Donald and Purver, Mark and Chapman, David and Fowler, Tom and Pouwels, Koen B. and Walker, A. Sarah and Peto, Tim E. A.},
  title   = {Effect of {Covid-19} vaccination on transmission of {Alpha} and {Delta} variants},
  journal = {New England Journal of Medicine},
  volume  = {386},
  number  = {8},
  pages   = {744--756},
  year    = {2022},
  doi     = {10.1056/NEJMoa2116597}
}

@article{hyndman2008forecast,
  author  = {Hyndman, Rob J. and Khandakar, Yeasmin},
  title   = {Automatic time series forecasting: the forecast package for {R}},
  journal = {Journal of Statistical Software},
  volume  = {27},
  number  = {3},
  pages   = {1--22},
  year    = {2008},
  doi     = {10.18637/jss.v027.i03}
}

@article{hyndman2002ets,
  author  = {Hyndman, Rob J. and Koehler, Anne B. and Snyder, Ralph D. and Grose, Simone},
  title   = {A state space framework for automatic forecasting using exponential smoothing methods},
  journal = {International Journal of Forecasting},
  volume  = {18},
  number  = {3},
  pages   = {439--454},
  year    = {2002},
  doi     = {10.1016/S0169-2070(01)00110-8}
}

@article{assimakopoulos2000theta,
  author  = {Assimakopoulos, Vassilis and Nikolopoulos, Konstantinos},
  title   = {The theta model: a decomposition approach to forecasting},
  journal = {International Journal of Forecasting},
  volume  = {16},
  number  = {4},
  pages   = {521--530},
  year    = {2000},
  doi     = {10.1016/S0169-2070(00)00066-2}
}

@misc{garza2022statsforecast,
  author       = {Garza, Federico and Mergenthaler Canseco, Max and Challu, Cristian and Olivares, Kin G.},
  title        = {{StatsForecast}: Lightning fast forecasting with statistical and econometric models},
  howpublished = {PyCon Salt Lake City, Utah, US. \url{https://github.com/Nixtla/statsforecast}},
  year         = {2022}
}

@misc{achiam2023gpt,
  author = {{OpenAI}},
  title  = {{GPT-4} Technical Report},
  note   = {arXiv:2303.08774},
  year   = {2023}
}

@article{ashokkumar2026large,
  title={Large language models can predict the results of social science experiments},
  author={Ashokkumar, Ashwini and Hewitt, Luke and Ghezae, Isaias and Willer, Robb},
  journal={Nature},
  pages={115--122},
  year={2026},
  doi={10.1038/s41586-026-10742-x},

}

@article{cui2025large,
  title={A large-scale replication of scenario-based experiments in psychology and management using large language models},
  author={Cui, Ziyan and Li, Ning and Zhou, Huaikang},
  journal={Nature Computational Science},
  volume={5},
  number={8},
  pages={627--634},
  year={2025},
 doi={10.1038/s43588-025-00840-7}
}

@article{cheung2025large,
  title={Large language models show amplified cognitive biases in moral decision-making},
  author={Cheung, Vanessa and Maier, Maximilian and Lieder, Falk},
  journal={Proceedings of the National Academy of Sciences},
  volume={122},
  number={25},
  pages={e2412015122},
  year={2025},
 doi={10.1073/pnas.2412015122}
}

@article{xu2026comparing,
  title={Comparing the algorithmic fidelity of large language models in predicting human decision making: a case study of vaccination choice},
  author={Xu, Shaochong and Zhang, Liyue and Jamison, Amelia M and Saiyed, Samee and Hou, Abe Bohan and Du, Hongru and Gardner, Lauren M},
  journal={npj Digital Public Health},
  volume={1},
  number={1},
  pages={20},
  year={2026},
doi={10.1038/s44482-026-00026-6}
}

@article{qu2024performance,
  title={Performance and biases of large language models in public opinion simulation},
  author={Qu, Yao and Wang, Jue},
  journal={Humanities and Social Sciences Communications},
  volume={11},
  number={1},
  pages={1095},
  year={2024},
  doi     = {10.1057/s41599-024-03609-x},

}

@misc{liu2026simulating,
  author = {Liu, Runzhou and Jong, Claire and Li, Haoyang and Cao, Yiming
            and Yao, Qing and Yamana, Teresa and Pei, Sen and Du, Hongru},
  title  = {Simulating population compliance with pandemic interventions using large language models},
  note   = {medRxiv, doi:10.64898/2026.05.12.26352942},
  year   = {2026}
}

@article{binz2023using,
  title={Using cognitive psychology to understand GPT-3},
  author={Binz, Marcel and Schulz, Eric},
  journal={Proceedings of the National Academy of Sciences},
  volume={120},
  number={6},
  pages={e2218523120},
  year={2023},
doi={10.1073/pnas.2218523120}
}

@article{xie2025using,
  title={Using large language models to categorize strategic situations and decipher motivations behind human behaviors},
  author={Xie, Yutong and Mei, Qiaozhu and Yuan, Walter and Jackson, Matthew O},
  journal={Proceedings of the National Academy of Sciences},
  volume={122},
  number={35},
  pages={e2512075122},
  year={2025},
doi={10.1073/pnas.2512075122}
}

@article{hu2025generative,
  title={Generative language models exhibit social identity biases},
  author={Hu, Tiancheng and Kyrychenko, Yara and Rathje, Steve and Collier, Nigel and Van Der Linden, Sander and Roozenbeek, Jon},
  journal={Nature Computational Science},
  volume={5},
  number={1},
  pages={65--75},
  year={2025},
doi={10.1038/s43588-024-00741-1}
}

@article{luo2025large,
title={Large language models surpass human experts in predicting neuroscience results},
 author={Luo, Xiaoliang and Rechardt, Akilles and Sun, Guangzhi and Nejad, Kevin K. and Y{\'a}{\~n}ez, Felipe and Yilmaz, Bati and Lee, Kangjoo and Cohen, Alexandra O. and Borghesani, Valentina and Pashkov, Anton and Marinazzo, Daniele and Nicholas, Jonathan and Salatiello, Alessandro and Sucholutsky, Ilia and Minervini, Pasquale and Razavi, Sepehr and Rocca, Roberta and Yusifov, Elkhan and Okalova, Tereza and Gu, Nianlong and Ferianc, Martin and Khona, Mikail and Patil, Kaustubh R. and Lee, Pui-Shee and Mata, Rui and Myers, Nicholas E. and Bizley, Jennifer K. and Musslick, Sebastian and Poyraz Bilgin, Isil and Niso, Guiomar and Ales, Justin M. and Gaebler, Michael and Ratan Murty, N. Apurva and Loued-Khenissi, Leyla and Behler, Anna and Hall, Chloe M. and Dafflon, Jessica and Bao, Sherry Dongqi and Love, Bradley C.},
  doi={10.1038/s41562-024-02046-9},
  journal={Nature human behaviour},
  volume={9},
  number={2},
  pages={305--315},
  year={2025},
}

@article{li2026theory,
  title   = {Theory-Informed Generative Agents for Human Mobility Modeling},
  author  = {Li, Haoyang and Liu, Runzhou and Li, Yao and Wesolowski, Amy and Pei, Sen and Du, Hongru},
  journal = {Research Square},
  year    = {2026},
  month   = feb,
  doi     = {10.21203/rs.3.rs-8902418/v1},
  url     = {https://doi.org/10.21203/rs.3.rs-8902418/v1},
  note    = {Preprint, Version 1}
}

@article{bedson2021review,
  title={A review and agenda for integrated disease models including social and behavioural factors},
  author={Bedson, Jamie and Skrip, Laura A and Pedi, Danielle and Abramowitz, Sharon and Carter, Simone and Jalloh, Mohamed F and Funk, Sebastian and Gobat, Nina and Giles-Vernick, Tamara and Chowell, Gerardo and de Almeida, Jo{\~a}o Rangel and Elessawi, Rania
   and Scarpino, Samuel V. and Hammond, Ross A. and Briand, Sylvie
   and Epstein, Joshua M. and H{\'e}bert-Dufresne, Laurent
   and Althouse, Benjamin M.},
  journal={Nature human behaviour},
  volume={5},
  doi = {10.1038/s41562-021-01136-2},
  number={7},
  pages={834--846},
  year={2021},

}

@article{funk2015nine,
  title={Nine challenges in incorporating the dynamics of behaviour in infectious diseases models},
  author={Funk, Sebastian and Bansal, Shweta and Bauch, Chris T and Eames, Ken TD and Edmunds, W John and Galvani, Alison P and Klepac, Petra},
  journal={Epidemics},
  volume={10},
  pages={21--25},
  year={2015},
  doi={10.1016/j.epidem.2014.09.005}

}

@article{fenichel2011adaptive,
  title={Adaptive human behavior in epidemiological models},
  author={Fenichel, Eli P and Castillo-Chavez, Carlos and Ceddia, M Graziano and Chowell, Gerardo and Parra, Paula A Gonzalez and Hickling, Graham J and Holloway, Garth and Horan, Richard and Morin, Benjamin and Perrings, Charles and Springborn, Michael and Velazquez, Leticia
   and Villalobos, Cristina},
  journal={Proceedings of the National Academy of Sciences},
  volume={108},
  number={15},
  pages={6306--6311},
  year={2011},
  doi= {10.1073/pnas.1011250108},

}

@article{bedford2019new,
  title={A new twenty-first century science for effective epidemic response},
  author={Bedford, Juliet and Farrar, Jeremy and Ihekweazu, Chikwe and Kang, Gagandeep and Koopmans, Marion and Nkengasong, John},
  journal={Nature},
  volume={575},
  number={7781},
  pages={130--136},
  year={2019},
doi={10.1038/s41586-019-1717-y}
}

@article{saad2023dynamics,
  title={Dynamics in a behavioral--epidemiological model for individual adherence to a nonpharmaceutical intervention},
  author={Saad-Roy, Chadi M and Traulsen, Arne},
  journal={Proceedings of the National Academy of Sciences},
  volume={120},
  number={44},
  pages={e2311584120},
  year={2023},
doi={10.1073/pnas.2311584120}
}

@article{martcheva2021effects,
  title={Effects of social-distancing on infectious disease dynamics: an evolutionary game theory and economic perspective},
  author={Martcheva, Maia and Tuncer, Necibe and Ngonghala, Calistus N},
  journal={Journal of Biological Dynamics},
  volume={15},
  number={1},
  pages={342--366},
  year={2021},
doi={10.1080/17513758.2021.1946177}
}

@article{haw2022optimizing,
title={Optimizing social and economic activity while containing {SARS-CoV-2} transmission using {DAEDALUS}},

  author={Haw, David J. and Forchini, Giovanni and Doohan, Patrick and Christen, Paula and Pianella, Matteo and Johnson, Robert and Bajaj, Sumali and Hogan, Alexandra B. and Winskill, Peter and Miraldo, Marisa and White, Peter J. and Ghani, Azra C. and Ferguson, Neil M. and Smith, Peter C. and Hauck, Katharina D.},
  doi={10.1038/s43588-022-00233-0},
  journal={Nature Computational Science},
  volume={2},
  number={4},
  pages={223--233},
  year={2022},

}

@article{pangallo2024unequal,
  title={The unequal effects of the health--economy trade-off during the COVID-19 pandemic},
   author={Pangallo, Marco and Aleta, Alberto and del Rio-Chanona, R. Maria and Pichler, Anton and Mart{\'\i}n-Corral, David and Chinazzi, Matteo and Lafond, Fran{\c c}ois and Ajelli, Marco and Moro, Esteban and Moreno, Yamir and Vespignani, Alessandro and Farmer, J. Doyne},
  doi={10.1038/s41562-023-01747-x},
  journal={Nature Human Behaviour},
  volume={8},
  number={2},
  pages={264--275},
  year={2024},

}

@article{du2025improving,
  title={Improving policy design and epidemic response using integrated models of economic choice and disease dynamics with behavioral feedback},
  author={Du, Hongru and Zahn, Matthew V and Loo, Sara L and Alleman, Tijs W and Truelove, Shaun and Patenaude, Bryan and Gardner, Lauren M and Papageorge, Nicholas and Hill, Alison L},
  journal={PLOS Computational Biology},
  volume={21},
  number={10},
  pages={e1013549},
  year={2025},
doi={10.1371/journal.pcbi.1013549}
}

@article{dobson2023balancing,
  title={Balancing economic and epidemiological interventions in the early stages of pathogen emergence},
  author={Dobson, Andy and Ricci, Cristiano and Boucekkine, Raouf and Gozzi, Fausto and Fabbri, Giorgio and Loch-Temzelides, Ted and Pascual, Mercedes},
  journal={Science Advances},
  volume={9},
  number={21},
  pages={eade6169},
  year={2023},
    doi={10.1126/sciadv.ade6169},

}

@article{ash2022disease,
  title={Disease-economy trade-offs under alternative epidemic control strategies},
  author={Ash, Thomas and Bento, Antonio M and Kaffine, Daniel and Rao, Akhil and Bento, Ana I},
  journal={Nature Communications},
  volume={13},
  number={1},
  pages={3319},
  year={2022},
doi={10.1038/s41467-022-30642-8}
}

@article{chang2021mobility,
  title={Mobility network models of COVID-19 explain inequities and inform reopening},
  author={Chang, Serina and Pierson, Emma and Koh, Pang Wei and Gerardin, Jaline and Redbird, Beth and Grusky, David and Leskovec, Jure},
  journal={Nature},
  volume={589},
  number={7840},
  pages={82--87},
  year={2021},
doi={10.1038/s41586-020-2923-3}
}

@article{nouvellet2021reduction,
  title={Reduction in mobility and COVID-19 transmission},
    author={Nouvellet, Pierre and Bhatia, Sangeeta and Cori, Anne and Ainslie, Kylie E. C. and Baguelin, Marc and Bhatt, Samir and Boonyasiri, Adhiratha and Brazeau, Nicholas F. and Cattarino, Lorenzo and Cooper, Laura V. and Coupland, Helen and Cucunuba, Zulma M. and Cuomo-Dannenburg, Gina and Dighe, Amy and Djaafara, Bimandra A. and Dorigatti, Ilaria and Eales, Oliver D. and van Elsland, Sabine L. and Nascimento, Fabricia F. and FitzJohn, Richard G. and Gaythorpe, Katy A. M. and Geidelberg, Lily and Green, William D. and Hamlet, Arran and Hauck, Katharina and Hinsley, Wes and Imai, Natsuko and Jeffrey, Benjamin and Knock, Edward and Laydon, Daniel J. and Lees, John A. and Mangal, Tara and Mellan, Thomas A. and Nedjati-Gilani, Gemma and Parag, Kris V. and Pons-Salort, Margarita and Ragonnet-Cronin, Manon and Riley, Steven and Unwin, H. Juliette T. and Verity, Robert and Vollmer, Michaela A. C. and Volz, Erik and Walker, Patrick G. T. and Walters, Caroline E. and Wang, Haowei and Watson, Oliver J. and Whittaker, Charles and Whittles, Lilith K. and Xi, Xiaoyue and Ferguson, Neil M. and Donnelly, Christl A.},
  doi={10.1038/s41467-021-21358-2},
  journal={Nature Communications},
  volume={12},
  number={1},
  pages={1090},
  year={2021},

}

@article{gimma2022changes,
  title={Changes in social contacts in England during the COVID-19 pandemic between March 2020 and March 2021 as measured by the CoMix survey: A repeated cross-sectional study},
  author={Gimma, Amy and Munday, James D and Wong, Kerry LM and Coletti, Pietro and van Zandvoort, Kevin and Prem, Kiesha and {{CMMID COVID-19 working group}} and Klepac, Petra and Rubin, G James and Funk, Sebastian and Edmunds, W. John and Jarvis, Christopher I.},
  journal={PLoS medicine},
  volume={19},
  number={3},
  pages={e1003907},
  year={2022},
  doi= {10.1371/journal.pmed.1003907},

}

@article{feehan2021quantifying,
  title={Quantifying population contact patterns in the United States during the COVID-19 pandemic},
  author={Feehan, Dennis M and Mahmud, Ayesha S},
  journal={Nature Communications},
  volume={12},
  number={1},
  pages={893},
  year={2021},
doi={10.1038/s41467-021-20990-2}
}

@article{jarvis2021impact,
  title={The impact of local and national restrictions in response to COVID-19 on social contacts in England: a longitudinal natural experiment.},
  author={Jarvis, Christopher I and Gimma, Amy and van Zandvoort, Kevin and Wong, Kerry LM and Edmunds, W John and {{CMMID COVID-19 working group}}},
  journal={BMC medicine},
  volume={19},
  number={1},
  pages={52},
  doi={10.1186/s12916-021-01924-7},
  year={2021},

}

@article{weitz2020awareness,
  title={Awareness-driven behavior changes can shift the shape of epidemics away from peaks and toward plateaus, shoulders, and oscillations},
  author={Weitz, Joshua S and Park, Sang Woo and Eksin, Ceyhun and Dushoff, Jonathan},
  journal={Proceedings of the National Academy of Sciences},
  volume={117},
  number={51},
  pages={32764--32771},
  year={2020},
doi={10.1073/pnas.2009911117}
}

@article{prem2021projecting,
  author = {
    Prem, Kiesha and van Zandvoort, Kevin and Klepac, Petra and Eggo, Rosalind M. and
    Davies, Nicholas G. and {{Centre for the Mathematical Modelling of Infectious Diseases COVID-19 Working Group}} and Cook, Alex R. and Jit, Mark
  },
  title = {
    Projecting contact matrices in 177 geographical regions:
    An update and comparison with empirical data for the {COVID-19} era
  },
  journal = {PLOS Computational Biology},
  year = {2021},
  volume = {17},
  number = {7},
  pages = {e1009098},
  doi = {10.1371/journal.pcbi.1009098},
  url = {https://doi.org/10.1371/journal.pcbi.1009098}
}

@article{dan2025surveyfatigue,
  author = {Dan, Shozen and Ling, Zhi and Chen, Yu and Tegegne, Joshua and Jaeger, Veronika K. and Karch, Andr{\'e} and Mishra, Swapnil and Ratmann, Oliver and
    {{Machine Learning \& Global Health network}}
  },
  title = {
Addressing survey fatigue bias in longitudinal social contact studies to improve pandemic preparedness
  },
  journal = {Scientific Reports},
  year = {2025},
  volume = {15},
  number = {1},
  pages = {17935},
  doi = {10.1038/s41598-025-02235-0},
  url = {https://doi.org/10.1038/s41598-025-02235-0}
}

@article{mao2026identifying,
  author = { Mao, Yicheng and Deardon, Rob and
    Deeth, Lorna E.
  },
  title = { Identifying memory mechanisms in Bayesian models of behavioural change during epidemics
  },
  journal = {Epidemics},
  year = {2026},
  volume = {56},
  pages = {100927},
  doi = {10.1016/j.epidem.2026.100927},
  url = {https://doi.org/10.1016/j.epidem.2026.100927},
  note = {Advance online publication}
}

@article{rikani2026resetting,
  title={Resetting population mobility responses under repeated nonpharmaceutical interventions: Implications for hypothesized pandemic fatigue},
  author={Rikani, Albano and Di Domenico, Laura and Sabbatini, Chiara E and Navarro, Victor and Ferres, Leo and Raude, Jocelyn and Colizza, Vittoria},
  journal={Proceedings of the National Academy of Sciences},
  volume={123},
  number={23},
  pages={e2533284123},
   doi={10.1073/pnas.2533284123},
  year={2026},

}

@article{funk2010modelling,
  title={Modelling the influence of human behaviour on the spread of infectious diseases: a review},
  author={Funk, Sebastian and Salath{\'e}, Marcel and Jansen, Vincent AA},
  journal={Journal of the Royal Society Interface},
  volume={7},
  number={50},
  pages={1247},
  year={2010}
}

@article{hamilton2024incorporating,
  title={Incorporating endogenous human behavior in models of COVID-19 transmission: A systematic scoping review},
  author={Hamilton, Alisa and Haghpanah, Fardad and Tulchinsky, Alexander and Kipshidze, Nodar and Poleon, Suprena and Lin, Gary and Du, Hongru and Gardner, Lauren and Klein, Eili},
  journal={Dialogues in Health},
  volume={4},
  pages={100179},
  year={2024},
  publisher={Elsevier}
}

@inproceedings{aher2023using,
  title={Using large language models to simulate multiple humans and replicate human subject studies},
  author={Aher, Gati V and Arriaga, Rosa I and Kalai, Adam Tauman},
  booktitle={International conference on machine learning},
  pages={337--371},
  year={2023},
  organization={PMLR}
}

@article{argyle2023out,
  title={Out of one, many: Using language models to simulate human samples},
  author={Argyle, Lisa P and Busby, Ethan C and Fulda, Nancy and Gubler, Joshua R and Rytting, Christopher and Wingate, David},
  journal={Political Analysis},
  volume={31},
  number={3},
  pages={337--351},
  year={2023},
  publisher={Cambridge University Press}
}

@article{gao2025take,
  title={Take caution in using LLMs as human surrogates},
  author={Gao, Yuan and Lee, Dokyun and Burtch, Gordon and Fazelpour, Sina},
  journal={Proceedings of the National Academy of Sciences},
  volume={122},
  number={24},
  pages={e2501660122},
  year={2025},
  publisher={National Academy of Sciences}
}

@article{xie2026evaluating,
  title={Evaluating the statistical realism of LLM-generated social science data},
  author={Xie, Yueqi and Liang, Lemeng and Li, Shuzhen and Lu, Yifu and Xiao, Zhiwen and Shi, Mengdi and Huang, Junming and Wang, Mengdi and Xie, Yu},
  journal={Proceedings of the National Academy of Sciences},
  volume={123},
  number={19},
  pages={e2538145123},
  year={2026},
  publisher={National Academy of Sciences}
}
\bibliographystyle{sciencemag}


\section*{Acknowledgments}


\paragraph*{Author contributions:} Y.M.: conceptualization, investigation, methodology, formal analysis, software, data curation, validation, visualization, resources, writing–original draft, and writing–review and editing. H.L.: validation, visualization, and writing–review and editing. R.D.: conceptualization; funding acquisition; supervision; writing–review and editing. H.D.: conceptualization, investigation, methodology, formal analysis, validation, resources, funding acquisition, supervision, writing–original draft, and writing–review and editing.

\paragraph*{Competing interests:} The authors declare no competing interests.






\newpage


\renewcommand{\thefigure}{S\arabic{figure}}
\renewcommand{\thetable}{S\arabic{table}}
\renewcommand{\theequation}{S\arabic{equation}}
\renewcommand{\thepage}{S\arabic{page}}
\setcounter{figure}{0}
\setcounter{table}{0}
\setcounter{equation}{0}
\setcounter{page}{1} 


\begin{center}
\section*{Supplementary Materials for\\ \scititle}

Yicheng~Mao,
Haoyang~Li,
Rob~Deardon,
Hongru~Du$^{\ast}$\\ 
\small$^\ast$Corresponding author. Email: hongrudu@virginia.edu\\

\end{center}

\subsubsection*{This PDF file includes:}
Materials and Methods\\
Supplementary Text\\
Figures S1 to S4\\
Tables S1 to S5\\

\newpage

\section*{Materials and Methods}

\subsection*{Data sources}
This section describes the data used to construct the behavioral inputs, generate contact matrices, calibrate the transmission model, and evaluate its outputs.

\paragraph{Pre-pandemic contact diaries.}
Contact behavior was taken from the French population-based contact survey
\cite{beraud2015french}, accessed through the SOCRATES data tool
\cite{willem2020socrates}. Each participant reports their own age, the day of the
week, whether the day was a holiday, and, for every person contacted, that person's
age and the setting of the contact among home, work, school, transport, leisure, and
other. We used the regular-weekday, non-holiday diaries and excluded participants who
reported more than twenty professional contacts, so that a small number of very
high-contact occupations does not dominate the average. After these restrictions, a
panel of 650 participants remained, made up of 211 children, 76 adolescents, 211
adults, and 152 seniors, each contributing one diary.
The number of participants in each age class is the denominator used in the
aggregation below under Contact-matrix construction. Each participant also reports the
number of people in their household, which sets the household floor used there.

\paragraph{Government control measures.}
Weekly policy was taken from the Oxford COVID-19 Government Response Tracker
\cite{hale2021oxcgrt} at the national level for France. We used seven ordinal
containment indicators, C1 through C7, covering school closing, workplace closing,
cancellation of public events, restrictions on gatherings, public transport,
stay-at-home requirements, and restrictions on internal movement. For each week we took the modal level of each indicator over the seven days of the week. The mapping from indicator levels to the English wording shown to the LLM is given below under LLM prompts.

\paragraph{Epidemic surveillance shown to the LLM.}
The weekly epidemic state used in retrospective reconstruction was built from nationally reported cases and deaths \cite{owid_covid}, aggregated to the week. Reported cases carry a reliability flag for the early period, when testing capacity was limited.
When cases were not reliably reported, the case counts were withheld from the prompt and only deaths were shown. Each week presents the current week and the previous week for comparison, together with the running cumulative totals.

\paragraph{Hospital admissions for calibration and prospective simulation.}
The transmission model was calibrated and scored against daily hospital admissions
from the SI-VIC database maintained by Sant\'e publique France \cite{spf_sivic},
resolved into the four age classes. The daily counts were corrected for notification
delays and missing days were imputed, following \cite{didomenico2026natcommun}.
Admissions summed over age groups were used as the calibration target. In the
retrospective contact reconstruction, hospital admissions were not shown to the
LLM. In the prospective forecasting, policy-conditioned, and
counterfactual analyses, admissions available before the target week formed the
epidemic input to the LLM and were subsequently extended using the
model's own preceding projections, as described below.

\subsection*{LLM prompts and queries}
For each participant and each week, the LLM receives one system message
and one user message. The system message is identical for all participants in a given
week and carries the task instruction and the situation of that week. The user
message carries the participant's age and diary. This section gives the templates in
full, the rules that build the weekly situation, and the way the queries were issued.

\paragraph{Task instruction.}
The system message opens with a fixed instruction that defines the task, the disease
background, and the six contact settings. The disease is described in general terms
and is not named. The full instruction is reproduced verbatim as the first part of the
worked example below.

\paragraph{Weekly situation.}
After the instruction, the system message appends the control policy in plain
language, the duration of the current policy configuration, and the epidemic
state. The message also identifies whether the target week falls within the
normal school term or an academic holiday and states how many weeks control
measures of some kind have been in force in total.

\paragraph{Policy wording.}
Each of the seven tracker indicators is turned into one line of plain English by the
level map below. For each indicator the modal weekly level selects one phrase.

\begin{quote}\small
\textbf{Schools:} 0, no school closing measures; 1, school closing recommended or
schools operating with modifications; 2, some categories of schools required to close;
3, all levels of schools required to close.\\
\textbf{Workplaces:} 0, no workplace closing measures; 1, workplace closing or working
from home recommended; 2, some sectors required to close or work from home; 3, all but
essential workplaces required to close.\\
\textbf{Public events:} 0, no public event restrictions; 1, cancellation recommended;
2, cancellation required.\\
\textbf{Gatherings:} 0, no restrictions; 1, above 1000 people; 2, 101 to 1000 people;
3, 11 to 100 people; 4, 10 people or fewer.\\
\textbf{Public transport:} 0, no measures; 1, closure recommended or substantially
reduced; 2, required to close or prohibited for most people.\\
\textbf{Stay-at-home:} 0, no requirement; 1, recommended not to leave home; 2, required
to stay home except for essential trips, exercise, and groceries; 3, required to stay
home with minimal exceptions.\\
\textbf{Internal movement:} 0, no restrictions; 1, between regions discouraged; 2,
restrictions in place.
\end{quote}

\paragraph{Policy duration.}
The message states how long the current set of measures has been unchanged. Starting
from the first lockdown week, we count back week by week while every one of the seven
policy lines matches the current week. If the run is one week, the message says the
measures came into force this week. Otherwise it states the number of consecutive
unchanged weeks. This exposes accumulating adherence fatigue to the model.

\paragraph{Epidemic state.}
In retrospective reconstruction, the epidemic block reports the current week's
new cases and deaths together with the previous week and the cumulative totals.
This block is concurrent with the week being generated. When case reporting is
flagged unreliable for the week, the case lines are replaced by a statement
that case counts are not reliably reported in that period, and only deaths are
shown.

In prospective settings, the epidemic block is instead constructed from
hospital admissions and reports the epidemic state through the previous week.
For weeks already projected beyond the origin, reported admissions are extended
using the model's own preceding predictions. The setting-specific construction
of this block is summarized in Table~\ref{tab:settings} and described below
under Real-time forecasting and Counterfactual policy experiments.

\paragraph{Academic-holiday adjustment.}
The prompt identifies whether schools are in session during the target week. During a normal teaching week, the message states that schools are in their normal teaching term. During an academic holiday, it states that no teaching is in session and that the closure follows the ordinary calendar rather than a control policy. The academic-holiday weeks were fixed as 6 July to 30 August 2020 (summer), 19 October to 1 November 2020 (autumn), 21 December 2020 to 3 January 2021 (Christmas), 8 February to 7 March 2021 (winter), and 12 to 25 April 2021 (spring). The same calendar was used in the deterministic matrix adjustment, where contacts occurring exclusively at school were assigned zero weight during academic holidays.

\paragraph{User message.}
The user message states the participant's age and lists their diary, one numbered line
per contact, each giving the contact's age and the setting. The model is asked which of
these contacts would still take place in the current week and to report their numbers.
The reply is a single JSON object of the form \verb|{"still_happen": [ ... ]}|. For participant $p$ in week $t$, the complete model reply can therefore be
represented by the binary retention vector
\begin{equation}
\mathbf{d}_p(t)
=
\left(
d_{p1}(t),\ldots,d_{pK_p}(t)
\right),
\qquad
d_{pk}(t)\in\{0,1\},
\end{equation}
where $K_p$ is the number of contacts in participant $p$'s baseline diary and
$d_{pk}(t)=1$ when contact $k$ appears in the \verb|still_happen| list. The
LLM evaluates the complete diary in a single query; the elements of
$\mathbf{d}_p(t)$ are the resulting contact-specific decisions.

\paragraph{Worked example.}
The blocks below reproduce one complete system message and user message exactly as sent to the model for a single participant in the analysis. The week falls in the second national lockdown of autumn 2020, and the participant is a 45-year-old adult.

\begin{systemprompt}
You are simulating how one real person behaved during an epidemic.

Background: a febrile respiratory infectious disease is spreading in France. It transmits through close human contact and can cause serious illness. The government responds with measures such as lockdowns, curfews, school and workplace closures, and limits on gatherings to reduce transmission.

The six contact settings, each corresponding to a physical location of contact:

- home: contacts at home, including the people you live with and any visitors
- work: contacts at the workplace
- school: contacts at school, college, or university
- transport: contacts on public transport
- leisure: contacts during leisure activities such as bars, restaurants, sports, cultural venues
- otherplace: contacts at other locations such as shops, services, and other public places

You will be given one person's age and their actual contact diary from a national survey carried out before the epidemic, on an ordinary working day. The diary lists every single person they had face-to-face or physical contact with on that day, with that person's age and where the contact took place. These are real contacts reported by a real respondent, not estimates.

Role-play that person under the conditions described below.

First work out, from the diary alone, what this person's ordinary day looks like. Are they at school, in work, retired, or at home? If they work, what kind of work is it, and could it be done from home? Who do they live with? The number of contacts they have in each setting, and the ages of those contacts, tell you most of this.

Then go through the diary one contact at a time and decide, for each one, whether that particular contact would still take place on a comparable day in the current week.

Decide each contact on its own merits. Some considerations push towards giving a contact up:

- the law currently forbids or restricts the activity
- the setting is crowded, indoors, or otherwise risky
- the epidemic is severe or getting worse this week

Other considerations push towards keeping a contact:

- the activity is essential or unavoidable for this person
- giving it up would cost them their income or their job
- someone depends on them for care
- restrictions have been in force for many weeks now, and people have grown tired of them
- compliance is never perfect, and some restricted contact always continues

Weigh both sides against this specific person and this specific contact. Model realized behavior, not perfect compliance, and do not assume the person obeys every rule.

One thing to keep in mind. Contacts at home include the people you live with, and you keep seeing them every day even under a strict lockdown.

Report the numbers of the contacts that would still take place. Any number you leave out is treated as a contact that no longer takes place. If none of them would still take place, report an empty list.

Respond ONLY with a JSON object, no other text:

{"still_happen": [list of contact numbers]}

============================================================

Current control policies in France:

- Schools: some categories of schools required to close
- Workplaces: all but essential workplaces required to close
- Public events: public events required to be canceled
- Gatherings: restrictions on gatherings of 10 people or fewer
- Stay-at-home: required not to leave home except for essential trips, exercise, and grocery shopping
- Public transport: no public transport measures
- Internal movement: internal movement restrictions in place

These measures came into force this week.

Epidemic situation this week:

- New confirmed cases this week: 303,116, last week: 332,505
- Cumulative confirmed cases: 1,643,952
- New deaths this week: 4,794, last week: 3,154
- Cumulative deaths: 33,509

School calendar: schools are in their normal teaching term.

Control measures of some kind have been in force for 33 weeks in total.
\end{systemprompt}

\begin{userprompt}
You are a 45 year old person in France.

Your own contact diary from before the epidemic, on an ordinary working day. You had face-to-face or physical contact with these 8 people:
1. a person aged about 25 to 30, met at a shop, a service, or another public place
2. a person aged about 25 to 30, met at work
3. a person aged about 30 to 35, met at work
4. a person aged about 40 to 45, met at a shop, a service, or another public place
5. a person aged about 50 to 55, met on public transport
6. a person aged about 25 to 30, met during a leisure activity
7. a person aged about 40 to 45, met at home
8. a person aged about 40 to 45, met at home

Which of these contacts would still take place on a comparable day this week? Report their numbers.
\end{userprompt}

\paragraph{Queries and caching.}
Queries were issued at temperature $0.3$ in JSON mode. Each reply is a JSON object holding a \verb|still_happen| list, and contact numbers outside the range of the diary were discarded. Every reply was cached on disk, and the cache is released with the code, so any matrix can be rebuilt without new queries.

\paragraph{Prompt variants across settings.} The task instruction, user message, academic-calendar treatment, and reply
format are shared across settings. Three fields vary, and Table~\ref{tab:settings} sets out how. In the retrospective
setting the policy block is the observed one for the week being generated and the epidemic block reports that same week's cases and deaths. In the forecasting setting the policy block is frozen at the origin, its duration counter is extended by the number of weeks past the origin, and the epidemic block reports hospital admissions up to the previous week. In the policy-conditioned setting the policy block and its duration counter are those of the target week, and the
epidemic block is as in forecasting. In a counterfactual setting the policy block is rendered from the scenario levels by the same map that renders the factual block, and the duration counter is recomputed on the counterfactual level sequence.

\subsection*{Contact-matrix construction}
A single procedure turns a set of weighted contacts into a $4\times4$ contact-rate
matrix. The pre-pandemic baseline matrix and every weekly LLM matrix pass
through this same procedure, and the only thing that changes between them is the weight
each reported contact carries. This section gives the procedure first, then the weights
that produce each matrix.

\paragraph{From the diaries to a contact-rate matrix.}
The panel is the set of participants defined above under Data sources, and
$n_i$ is the number of participants in age class $i$. Each participant $p$
reports a list of contacts, and contact $k$ has age class $g(k)$, taken from
its reported exact age when available and otherwise from the midpoint of its
reported age range. Each contact enters the sum below with a weight
$w^{(p)}_k$ between zero and one.

Writing $a(p)$ for the participant's age class, the mean number of contacts
reported by a participant in age class $i$ with people in age class $j$ is
\begin{equation}
M_{ij}
=
\frac{1}{n_i}
\sum_{p:\,a(p)=i}
\sum_{k:\,g(k)=j}
w^{(p)}_k .
\end{equation}
Thus, rows of $M$ index participant age and columns index contact age.

Reported contacts are not symmetric between age classes because the two
classes are sampled at different rates. Using population sizes $N_i$, we
applied the reciprocity correction
\begin{equation}
M^{\mathrm r}_{ij}
=
\frac{M_{ij}N_i+M_{ji}N_j}{2N_i}.
\end{equation}
The reciprocity-corrected matrix retains the same orientation: row $i$
represents participant age and column $j$ represents contact age.

The matrix supplied to the transmission model was
\begin{equation}
C_{ij}
=
\frac{M^{\mathrm r}_{ij}}{N_j}N,
\qquad
N=\sum_k N_k.
\end{equation}
Here $C_{ij}$ is the rate at which a member of participant age group $i$
contacts members of age group $j$.

For visualization and structural comparison, we used the transposed display
matrix
\begin{equation}
D_{ij}
=
M^{\mathrm r}_{ji},
\label{eq:display_matrix}
\end{equation}
so that rows of $D$ index contact age and columns index participant age.
All contact-matrix heatmaps and the supplementary structural comparison use
this display orientation.

\paragraph{The pre-pandemic baseline matrix.}
Setting $w^{(p)}_k = 1$ for every contact reported in the panel diaries and applying the
three equations above gives the pre-pandemic baseline matrix. These weights and these
equations are its complete definition, so the baseline follows from the survey alone.

\paragraph{Weights for the LLM matrices.}
A weekly LLM matrix uses the same reconstruction equations with
weights derived from the binary retention vector $\mathbf{d}_p(t)$. Before
deterministic adjustments, contact $k$ of participant $p$ has weight
\begin{equation}
w_k^{(p)}(t)=d_{pk}(t).
\end{equation}
The school-calendar and household rules described below then modify these
initial weights where required. The resulting $w_k^{(p)}(t)\in[0,1]$ are the
final weights entering the matrix reconstruction.

The first is the school calendar. On an academic-holiday week, any contact whose only
setting is school is set to zero, since schools are not in session and such a contact
cannot occur whatever the model decides. The holiday weeks are the fixed set listed
above under LLM prompts.

The second is a floor on contact within the household. Let $n_{\mathrm{home}}$ be the
number of household members the participant reported and let $s$ be the sum of the
home-contact weights after the model's decision. If $s$ falls below
\begin{equation}
f = \min\!\left(\bar{h}-1,\; n_{\mathrm{home}}\right),\qquad \bar{h}=2.18,
\end{equation}
the home-contact weights are raised evenly, each capped at one, until they sum to $f$.
Here $\bar{h}=2.18$ is the average household size in France \cite{insee_households}, so
$\bar{h}-1$ is the average number of co-residents. The floor keeps household contact from
being removed entirely, since co-residents keep meeting even under a strict stay-at-home
order. It is skipped when the participant reports no household contact.

\paragraph{Contact intensity.}
For participant age group $i$, age-specific contact intensity was defined as
\begin{equation}
c_i
=
\sum_j C_{ij}\frac{N_j}{N}
=
\sum_j M^{\mathrm r}_{ij}.
\label{eq:participant_contact_intensity}
\end{equation}
This is the mean daily number of contacts reported by a member of participant
age group $i$ after reciprocity correction. Equivalently, using the display
matrix defined in Eq.~\ref{eq:display_matrix},
\begin{equation}
c_i
=
\sum_j D_{ji},
\end{equation}
that is, the sum of column $i$ of the displayed matrix.

Aggregate contact intensity was the population-weighted average
\begin{equation}
\bar{c}
=
\sum_i \frac{N_i}{N}c_i.
\end{equation}
The same definitions were applied to every matrix configuration and to all
retrospective, forecasting, and counterfactual analyses.

\subsection*{Transmission model}
Transmission follows the age-stratified stochastic model of
\cite{didomenico2020idf,didomenico2026natcommun}, run in the two-strain configuration for
the Wuhan-like strain and the Alpha variant. That work describes the model in full. This
section gives the structure and the parameter values used here.
 
\begin{figure}[p]
\centering
\includegraphics[
  width=\textwidth,
  height=0.60\textheight,
  keepaspectratio
]{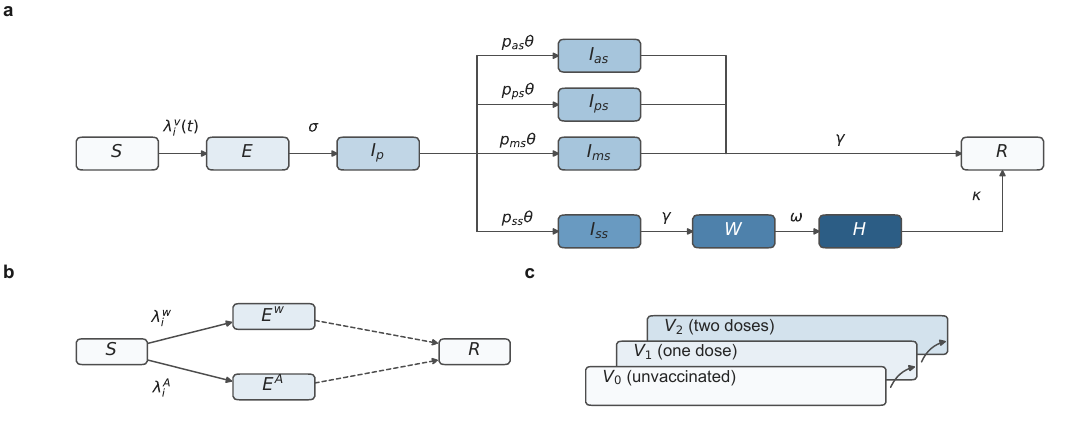}
\caption{\textbf{Structure of the transmission model.}
(\textbf{a}) Compartments and transitions within one age class, one viral strain,
and one vaccination stratum. Susceptible individuals ($S$) acquire strain $v$
at rate $\lambda_i^v(t)$ and enter the latent state ($E$). Individuals progress from the
presymptomatic state ($I_p$) to asymptomatic ($I_{as}$), pauci-symptomatic
($I_{ps}$), mildly symptomatic ($I_{ms}$), or severely symptomatic ($I_{ss}$)
infection according to age- and strain-specific probabilities. Only severe
cases enter the hospital pathway, passing through a waiting state ($W$) that
represents the delay between symptom onset and hospital admission. The
transition from $W$ to $H$ corresponds to the daily hospital admissions used
for model calibration. Rate symbols and parameter values are given in
Table~\ref{tab:nat}. (\textbf{b}) Strain strata. Susceptible individuals acquire
either the Wuhan-like strain ($w$) or the Alpha variant ($A$), progress through
the compartment structure shown in (\textbf{a}) (dashed arrows), and share a
common recovered state because recovery from either strain confers
cross-immunity. (\textbf{c}) Vaccination strata. The compartment structure in
(\textbf{a}) is replicated across vaccination levels. Transitions between dose
levels follow the observed vaccination schedule for adults and seniors.
Vaccination modifies susceptibility, transmissibility, symptom probability, and
disease severity (Table~\ref{tab:vax}).}
\label{fig:model_structure}
\end{figure}

\paragraph{Compartments.}
Each age class holds susceptible ($S$), latent ($E$), and presymptomatic infectious
($I_p$) individuals, then one asymptomatic infectious state ($I_{as}$) and three
symptomatic infectious states of increasing severity, pauci-symptomatic ($I_{ps}$),
mildly symptomatic ($I_{ms}$), and severely symptomatic ($I_{ss}$). A severe case moves
to a waiting state ($W$), then to hospital admission ($H$), and then to recovery ($R$).
Daily hospital admissions, the calibration target used throughout, are the daily flow from $W$ into $H$.
The exposed, infectious, waiting, and hospitalized compartments are stratified by strain $v$, taking the values $w$ for the Wuhan-like strain and $A$ for the Alpha variant. Susceptible individuals are shared across strains, and recovery from either strain enters a common recovered state. The compartment structure is additionally stratified by vaccination status among unvaccinated ($V_0$), one dose ($V_1$), and two doses ($V_2$).
The compartments and the two strata are shown in Fig.~\ref{fig:model_structure}.
 
\paragraph{Force of infection.}
Let $C_{ij}(t)$ denote the population contact matrix supplied by the
behavioral layer. For infectious compartment $c$, symptom- and
testing-associated contact reductions described below map this matrix to
$C^c_{ij}(t)$. Let $\beta$ be the per-contact transmission rate and $N$ the
total population. The rate at which a susceptible individual in age class $i$
acquires strain $v$ is
\begin{equation}
\lambda^{v}_i(t) = \beta\,\alpha(t)\,\eta^{v} s^{v}_i \sum_{c}\sum_{j}
r_{cj}\, C^{c}_{ij}(t)\, \frac{I^{v}_{cj}(t)}{N},
\end{equation}
where $c$ runs over the five infectious states, $s^{v}_i$ is the age- and strain-specific
susceptibility, $\eta^{v}$ the transmission advantage of the strain, $r_{cj}$ the relative
infectiousness of age class $j$ in compartment $c$, and $I^{v}_{cj}(t)$ the corresponding
infectious count. The correcting factor $\alpha(t)$ scales the overall level of contact
and leaves the relative pattern across ages fixed. It is piecewise constant over the
fitting windows and is estimated as described below under Model inference. The transmission-model structure and all epidemiological inputs are held fixed
across matrix configurations. 
Presymptomatic, asymptomatic, and pauci-symptomatic cases carry the reduced infectiousness given in Table~\ref{tab:nat}. Vaccination lowers $s^{v}_i$ by
the vaccine effect against infection, and it lowers the contribution of a vaccinated
infectious case by the effect against onward transmission. All contacts contribute
equally to transmission, with no distinction by setting or by the physical nature of the
contact.
 
\paragraph{Contacts by disease stage.}
The behavioral-layer matrix $C_{ij}(t)$ is modified within the transmission
model to obtain the compartment-specific contact rate $C^c_{ij}(t)$. Two disease-related adjustments are applied. 
A case with severe
symptoms reduces its contacts by 75 percent, which represents spontaneous isolation
during illness. A fraction of infectious individuals is identified by testing and reduces
its contacts by 90 percent. That fraction is zero through the first wave and the first
lockdown, when systematic testing was not yet in place in France, and 50 percent from the
exit of the first lockdown onwards. Presymptomatic cases keep their contacts, which
accounts for the delay between infection and a positive test.
 
\paragraph{Progression and severity.}
States are left at constant per-day rates set by the mean durations in
Table~\ref{tab:nat}. From $I_p$ a case branches into the asymptomatic state and the three
symptomatic states with age- and strain-specific probabilities. The probability
of mild symptoms is the remaining probability,
\begin{equation}
p_{ms}
=
1-p_{as}-p_{ps}-p_{ss}.
\end{equation}
The probability of severe symptoms rises steeply with age and is raised for
Alpha by the hospitalization multiplier in Table~\ref{tab:nat}. 
Only severe cases reach the hospital pathway.

\paragraph{Strains and vaccination.}
The Alpha variant is introduced during the winter of 2020 to 2021 at the prevalence
observed in variant surveillance \cite{gaymard2021alpha}, carrying the transmission
advantage and the raised hospitalization probability of Table~\ref{tab:nat}. Vaccination
applies to adults and seniors, moving them through one-dose and two-dose states at the
pace of the administered-dose record \cite{spf_vacsi}, up to a maximum coverage of
$0.99$. Each dose level carries four effects (Table~\ref{tab:vax}). The effects against
infection and against onward transmission enter the force of infection, lowering the
susceptibility $s^{v}_i$ and the infectiousness of vaccinated cases, respectively. The
effects against symptomatic and against severe disease instead act on disease
progression, reducing for vaccinated individuals the probability of developing
symptomatic rather than asymptomatic or pauci-symptomatic infection and, among
symptomatic cases, the probability of severe disease. These latter two effects are
applied to the probability of disease given infection, so that the protection against
infection already represented in the force of infection is not counted twice.

\paragraph{Integration.}
The epidemic is seeded with ten presymptomatic adults on the first day of the simulation, taken as a fixed start date of 16 January 2020. The model is advanced by $\tau$-leaping with a one-day step, with
state transitions drawn from binomial and multinomial distributions. Each configuration is run as an ensemble,
and trajectories are summarized by the median and the 2.5th and 97.5th percentiles.

\subsection*{Model inference}
\paragraph{Transmission rate.}
The per-contact transmission rate $\beta$ is fitted in the pre-lockdown phase, before
any control measure, with contacts held at the pre-pandemic baseline. We evaluated the simulation-based Poisson objective defined below over a grid
of $\beta$ from $0.060$ to $0.120$ in steps of $0.002$, using 50 stochastic
runs at each grid point, and selected the maximizing value. This gives $\beta = 0.098$. Because every retrospective configuration uses the pre-pandemic baseline matrix during
the pre-lockdown period, this single value of $\beta$ is shared across configurations.
 
\paragraph{Correcting factors.}
With $\beta$ fixed, one value of $\alpha$ is fitted for each window in turn, starting
with the window that carries the lockdown matrix. Each value is searched on the interval
$[0.15, 3.0]$ by golden-section to a width of $0.3$, then refined by a quadratic fit to
the log-likelihood on a grid of width $0.2$ in steps of $0.02$. When the optimum sits at
a search boundary, the interval is widened and the search repeats. The coarse
golden-section stage uses 10 stochastic runs and the refinement uses 50.
 
\paragraph{Simulation-based calibration objective.}
For a fitting window $[t_1,t_2]$, let $y_t$ denote observed daily hospital
admissions summed over age groups. For each candidate parameter value, the
transmission model was simulated repeatedly and the daily ensemble median
$\tilde{\mu}_t(\Theta)$ was calculated. Candidate values were compared using
the simulation-based Poisson objective
\begin{equation}
\ell(\Theta)
=
\sum_{t=t_1}^{t_2}
\log
\mathrm{Poisson}
\left(
y_t;\tilde{\mu}_t(\Theta)
\right).
\end{equation}
The ensemble median was used to reduce the influence of highly variable
stochastic realizations during numerical calibration. This objective was used
for parameter selection and was not interpreted as a full probabilistic
likelihood for the stochastic transmission process.
 
\paragraph{Fitting windows.}
In the retrospective fit the lockdown window covers 16 March to 10 May 2020,
and the period from 11 May 2020 to 4 July 2021 is fitted in fifteen
consecutive windows beginning on 11 May, 6 July, 31 August, 7 September,
5 October, 2 November, 23 November, and 21 December 2020, and on 18 January,
1 February, 15 February, 8 March, 22 March, 5 April, and 3 May 2021. Each
window ends on the Sunday before the next begins, and the last window ends on 4 July 2021. These track the exit from the first lockdown, the summer and school holidays, the second lockdown, the winter holidays, and the spring of 2021. The boundaries are those of \cite{didomenico2026natcommun}, so all five retrospective configurations are calibrated on the same partition of time as
the published mobility-driven matrix results and the comparison is not affected by the choice of windows. In the forecasting analysis each week from the week beginning 11 May 2020 to the forecast origin is its own window.

\subsection*{Real-time forecasting}
\paragraph{Origins and horizon.}
Forecasts are issued from rolling origins at the weeks beginning 14 September to 7 December 2020, covering the autumn second wave,
each projecting a four-week horizon. Each origin uses 100 stochastic runs and produces daily age-resolved quantile trajectories.
 
\paragraph{Real-time matrices.}
For each future week within the horizon, the LLM regenerates the
contact matrix without using observations obtained after the forecast origin. 
The policy is frozen at the origin, and its
duration counter is extended by the number of weeks past the origin. The school calendar
is published ahead of the term, so the calendar line for a future week is used as it
stands. The epidemic block
is built from hospital admissions and reports the state up to the previous week, so the
week being generated is never shown its own admissions. For weeks up to the origin these
admissions are the reported values, and for weeks already forecast within the horizon
they are the model's own median predictions. No observation past the origin enters the
prompt. This block lags the epidemic signal by one week relative to the retrospective
generation, where the concurrent week's cases and deaths are shown.
 
\paragraph{Frozen correcting factors.}
Weekly correcting factors are fitted up to the origin. Across the horizon they are frozen
forward at the mean of the last $K=4$ fitted weekly factors,
\begin{equation}
\alpha_{\mathrm{fwd}}(o) = \frac{1}{K}\sum_{w=o-K+1}^{o} \alpha_w .
\end{equation}
Weeks up to the origin keep their own fitted weekly values, and the lockdown period keeps
the factor fitted for 16 March to 10 May 2020.
 
\paragraph{Mobility-driven configuration.}
The mobility-driven configuration is run in the same real-time configuration. Its published weekly
matrices up to the origin are used unchanged, and its future-week matrices are frozen at the origin, so neither time-varying configuration
uses behavioral observations obtained after the forecast origin.

\paragraph{Weighted interval score.}
Probabilistic forecast accuracy was summarized by the weighted interval score
(WIS)~\cite{bracher2021wis}. For an observed target $y$, predictive median $m$,
and $K$ central prediction intervals with nominal levels $1-\alpha_k$ and bounds
$l_k$ and $u_k$,
\begin{equation}
\mathrm{WIS}
=
\frac{1}{K+1/2}
\left[
\frac{|y-m|}{2}
+
\sum_{k=1}^{K}
\left\{
\frac{\alpha_k}{2}(u_k-l_k)
+
(l_k-y)\mathbf{1}\{y<l_k\}
+
(y-u_k)\mathbf{1}\{y>u_k\}
\right\}
\right].
\label{eq:wis}
\end{equation}
Following the collaborative forecasting hubs~\cite{bracher2021wis}, we used the
predictive median together with $K=11$ central prediction intervals, with
$\alpha_k \in \{0.02, 0.05, 0.10, 0.20, 0.30, 0.40, 0.50, 0.60, 0.70, 0.80,
0.90\}$, corresponding to the standard set of 23 quantile levels.
 
\subsection*{Statistical forecast baselines}
Three purely statistical forecasters provide references for the relative score, with no
epidemiological structure. Automatic ARIMA selects the model order automatically
\cite{hyndman2008forecast}, automatic exponential smoothing selects an ETS model
automatically \cite{hyndman2002ets}, and the Theta method decomposes and extrapolates
the series \cite{assimakopoulos2000theta}. All three run through the statsforecast package
\cite{garza2022statsforecast} in a non-seasonal setting.
 
Each forecaster is fitted in real time. At an origin it uses only the weekly admissions
from the week beginning 11 May 2020 up to that origin, taken as complete Monday to Sunday sums, on the
$\log(1+x)$ scale so that forecasts stay non-negative, and produces the four-week
horizon. For the age-specific evaluation, each statistical forecaster was fitted independently to each age group's weekly admission series in the same way, using that group's admissions from the week beginning 11 May 2020 to the origin.
The central prediction
intervals are converted to the 23 quantile levels used by collaborative forecasting hubs,
clipped to be non-negative and non-decreasing. The reference set for the relative weighted
interval score is these three forecasters together with the mobility-driven matrix configuration.

\subsection*{Counterfactual policy experiments}

\paragraph{Analytical design.}
Each counterfactual experiment branches from a fixed origin week and projects
the following four weeks. Weeks up to and including the branch retain their
factual configuration. Beginning in the following week, the seven containment
indicators supplied to the LLM follow a prespecified scenario path.

All scenarios at a branch point share the epidemic state and contact matrices
up to the branch, the transmission parameters, vaccination and variant inputs,
the correcting factor frozen forward from the branch, and the random seeds.
The school calendar is also shared and continues to follow the observed
academic calendar. The scenario-specific policy sequence and its associated
duration counter are therefore the inputs that differ initially across
counterfactual runs. Subsequent epidemic-state prompts also diverge as each
scenario incorporates its own preceding admission projections.

\paragraph{Scenario stringency.}
Each scenario was placed on the Oxford stringency-index scale so that its
overall policy intensity could be compared with the factual scenario and with
the other scenarios. The index averages nine normalized sub-indices: the eight
containment indicators C1 to C8 and the public-information indicator H1
\cite{hale2021oxcgrt}.

For indicator $j$ in week $t$, let $v_{j,t}$ denote the recorded policy level,
$N_j$ its maximum possible level, $F_j$ indicate whether the indicator includes
a geographic-scope flag, and $f_{j,t}$ denote the corresponding flag value. The
normalized indicator score was
\begin{equation}
I_{j,t}
=
100
\frac{
v_{j,t}
-
0.5\left(F_j-f_{j,t}\right)
}{
N_j
},
\end{equation}
with the flag adjustment set to zero when $v_{j,t}=0$. The weekly stringency
index was the arithmetic mean of the nine normalized scores.

For a counterfactual week, indicators C1 to C7 took the levels specified by the
scenario. The international-travel indicator C8 and public-information
indicator H1 retained their observed values for that week. When a scenario
changed one of C1 to C7 relative to the observed value, its geographic-scope
flag was set to national coverage. Flags for unchanged indicators retained
their observed values. Each scenario was summarized by the mean of its four
weekly stringency-index values over the projection window.

\paragraph{Selection of branch points.}
Branch points were selected from the 13 rolling origins used in the forecasting
analysis. We sought one origin followed by a consequential tightening and one
followed by a consequential relaxation.

For an origin $o$, let $t=o+4$ denote the four-week target. Let
$m^{\mathrm{pol}}_t$ denote median projected admissions in target week $t$ when
the factual policy level of each projected week is supplied, and let
$m^{\mathrm{frz}}_t$ denote the corresponding projection when the policy levels
at origin $o$ are held throughout the horizon. With $y_t$ denoting observed admissions in the target week, the estimated
effect of the policy change was
\begin{equation}
\Delta_{\mathrm{pol}}(o)
=
\frac{
m^{\mathrm{frz}}_t
-
m^{\mathrm{pol}}_t
}{
y_t
}.
\end{equation}
Positive values of $\Delta_{\mathrm{pol}}(o)$ indicate that the factual policy
change reduced projected admissions relative to holding the origin policy
fixed, whereas negative values indicate that the factual change increased
projected admissions.

Table~\ref{tab:cf_cand} reports all candidate origins. The week beginning 26 October 2020 had the largest positive value and was selected as the lockdown branch. The week beginning 23 November 2020 had the most negative value and was selected as the relaxation branch.

\paragraph{Scenario specification.}
Six policy scenarios were evaluated at each branch point
(Table~\ref{tab:cf_scen}).

\begin{itemize}
\item \textit{Factual}: the containment levels subsequently observed in each
week.

\item \textit{No change}: the levels in force at the branch remain in place
throughout the four-week window.

\item \textit{Delayed by one week}: the factual policy sequence is shifted
forward by one week, with the branch-week levels retained until the delayed
change begins.

\item \textit{Delayed by two weeks}: the factual policy sequence is shifted
forward by two weeks.

\item \textit{Maximum measures}: all seven containment indicators are set to
their maximum levels throughout the projection window.

\item \textit{Schools closed}: the school-closing indicator is set to its
maximum level, while the other six containment indicators follow the factual
sequence.
\end{itemize}

At the week-44 branch, the factual scenario imposes a simultaneous tightening in the week beginning 2 November 2020. The delayed scenarios therefore postpone this tightening by one or two
weeks. At the relaxation branch, the factual scenario relaxes workplace restrictions in the week beginning 30 November 2020 and internal-movement restrictions in the week beginning 14 December 2020. The delayed scenarios shift both stages of this relaxation forward together.

For every scenario, the policy-duration counter was recomputed along the
scenario-specific sequence. The epidemic block was updated recursively using
reported admissions available at the branch and the model's own median
admission predictions for earlier weeks within the counterfactual horizon. No
hospital admission observed after the branch entered scenario generation.

\paragraph{Counterfactual outcome summaries.}
For each scenario, aggregate contact intensity was calculated for each projected
week and averaged over the four-week window. Age-specific contact intensities
were averaged over the same weeks. Hospital admissions were summed over the
four-week window. Changes in each quantity were expressed as percentages
relative to the factual-policy projection at the same branch point:
\begin{equation}
\Delta Q_s
=
100
\left(
\frac{Q_s}{Q_{\mathrm{factual}}}
-
1
\right),
\end{equation}
where $Q_s$ denotes the scenario-specific four-week summary of the quantity of interest.

We examined whether ordering the six scenarios by their mean four-week
stringency index produced corresponding orderings of aggregate contact
intensity and cumulative admissions. The association between behavioral and
epidemic responses was calculated across the ten non-factual scenarios using
the Pearson correlation between the percentage change in aggregate contact
intensity and the percentage change in cumulative admissions.

\paragraph{School-contact shares and age-specific responses.}
The age-specific response to school closure was compared with the composition
of contacts in the pre-pandemic diaries. Let $q^{\mathrm{school}}_p$ denote the
number of school contacts reported by respondent $p$, and let
$q^{\mathrm{all}}_p$ denote that respondent's total number of contacts. For age
group $i$, containing $n_i$ respondents, the school-contact share was defined as
\begin{equation}
s_i
=
\frac{
n_i^{-1}\sum_{p:\,a(p)=i} q^{\mathrm{school}}_p
}{
n_i^{-1}\sum_{p:\,a(p)=i} q^{\mathrm{all}}_p
}
=
\frac{
\sum_{p:\,a(p)=i} q^{\mathrm{school}}_p
}{
\sum_{p:\,a(p)=i} q^{\mathrm{all}}_p
}.
\end{equation}
Thus, the share is the mean number of school contacts per respondent divided by
the mean total number of contacts per respondent within the age group. 

\section*{Supplementary Text}

\subsection*{Structural comparison of LLM and mobility-driven contact matrices}

We compared each weekly LLM contact matrix with the corresponding
mobility-driven synthetic matrix over the weeks available for all four
time-varying configurations. Analyses covered the 68 weeks from 16 March 2020 to 4 July 2021, the period for which all four time-varying configurations were available. For this analysis, matrices were expressed in the display orientation used throughout the figures, with contact age indexing rows and participant age indexing columns.

To characterize the complete $4\times4$ matrix while distinguishing contacts
among older age groups from those involving younger age groups, we partitioned
the 16 matrix elements into two complementary sets. The first set comprised the
four elements in the adult--senior block, for which both participant and contact
age belonged to the adult or senior groups. The second set comprised the
remaining 12 elements, each involving at least one child or adolescent group.
Together, the two sets covered all elements of the contact matrix.

Let $\widetilde{M}^{(m)}_{ij,t}$ denote the reciprocity-corrected matrix element
for LLM $m$ in week $t$, where row $i$ indexes contact age and column
$j$ indexes participant age. For each LLM and each element set, we
estimated the slope of a linear regression constrained through the origin,
\begin{equation}
\widetilde{M}^{(m)}_{ij,t}
=
b_m
\widetilde{M}^{(\mathrm{mob})}_{ij,t}
+
\varepsilon_{ij,t},
\label{eq:structural_slope}
\end{equation}
where values of $b_m$ below and above one indicate systematically lower and
higher matrix-element magnitudes, respectively, relative to the mobility-driven matrices. We also calculated the Pearson correlation coefficient across matched
elements and weeks to quantify preservation of their relative ordering. The
regression slopes and correlations were calculated on the original matrix
values; logarithmic axes were used only for visualization.

For each of the four participant age groups, the within-group contact fraction
was calculated as
\begin{equation}
A^{(m)}_{j,t}
=
\frac{\widetilde{M}^{(m)}_{jj,t}}
     {\sum_i \widetilde{M}^{(m)}_{ij,t}},
\label{eq:within_group_fraction}
\end{equation}
where the denominator is the total contact intensity reported by participants
in age group $j$. Age-specific contact intensity was therefore defined as
\begin{equation}
c^{(m)}_{j,t}
=
\sum_i \widetilde{M}^{(m)}_{ij,t}.
\label{eq:age_contact_intensity}
\end{equation}
We then calculated the weekly contact-intensity ratio relative to the mobility-driven matrix,
\begin{equation}
R^{(m)}_{j,t}
=
\frac{c^{(m)}_{j,t}}
     {c^{(\mathrm{mob})}_{j,t}},
\label{eq:contact_ratio}
\end{equation}
and summarized this quantity by its median across matched weeks. Values below
one indicate lower age-specific contact intensity than in the mobility-driven matrices, whereas values above one indicate higher intensity.

\begin{figure}[p]
\centering
\includegraphics[
  width=\textwidth,
  height=0.7\textheight,
  keepaspectratio
]{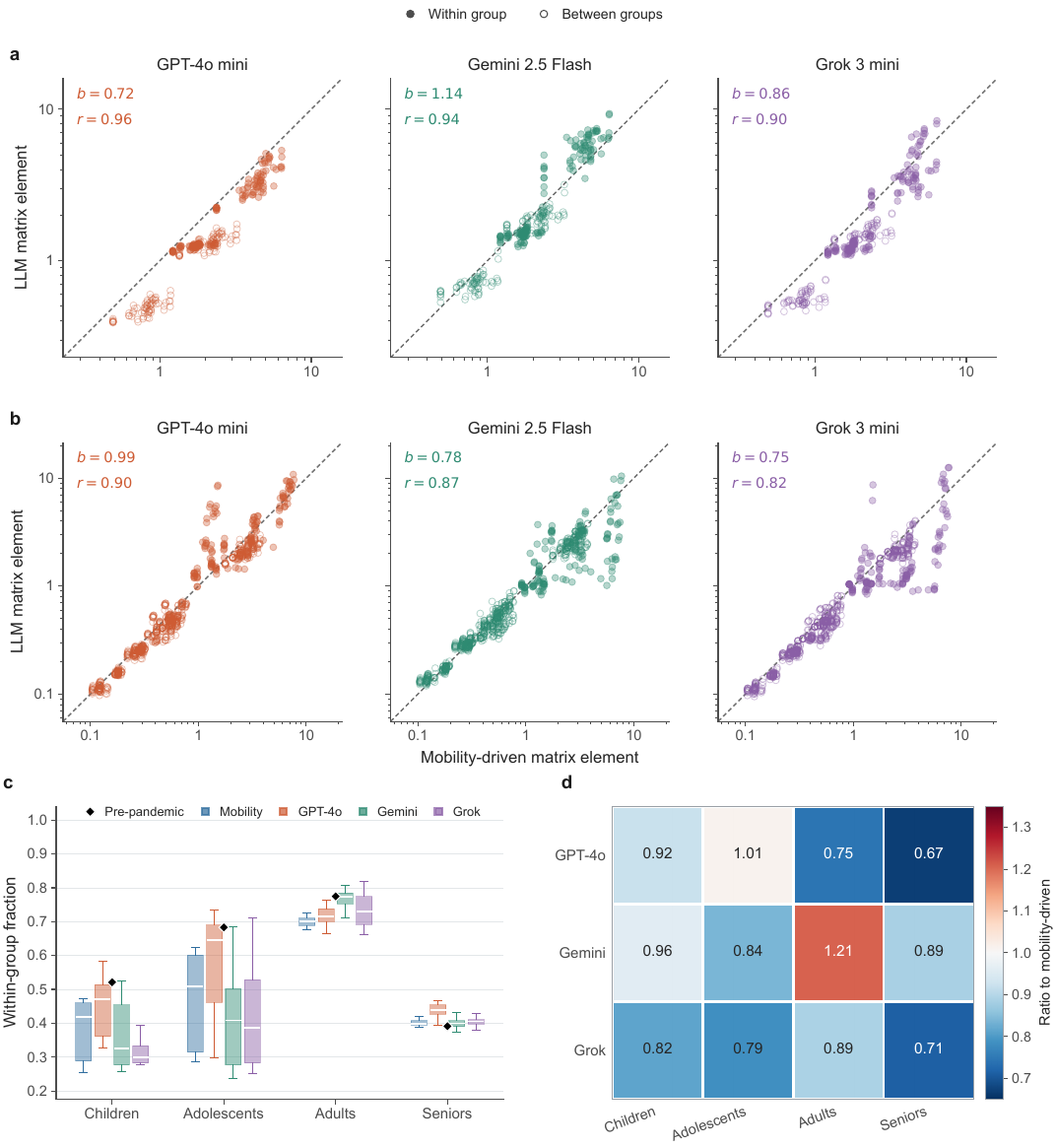}
\caption{\textbf{Structural comparison of LLM and mobility-driven contact
matrices.}
(\textbf{a}) Element-wise comparisons of the adult--senior matrix block for
GPT-4o mini, Gemini 2.5 Flash, and Grok 3 mini. Each point represents one
reciprocity-corrected matrix element in one matched week. Filled and open
symbols denote within- and between-group contacts, respectively. Both axes are
logarithmic; dashed lines indicate equality with the mobility-driven matrix.
Annotations report slopes from regressions constrained through the origin
($b$) and Pearson correlations ($r$).
(\textbf{b}) Corresponding comparisons for matrix elements involving children
or adolescents.
(\textbf{c}) Weekly distributions of the within-group contact fraction by
participant age group and matrix configuration. Boxes span the interquartile range,
center lines denote medians, and whiskers extend to 1.5 times the interquartile
range. Diamonds show the static pre-pandemic values.
(\textbf{d}) Median age-specific contact intensity relative to the mobility-driven matrices across matched weeks. Values below and above one indicate lower and
higher contact intensity, respectively.}
\label{fig:matrix_structure}
\end{figure}

GPT-4o mini generated systematically lower values in the adult--senior block,
with a regression slope of $b=0.72$, while strongly preserving the relative
ordering of these elements ($r=0.96$; Fig.~\ref{fig:matrix_structure}a).
Median contact intensity relative to the mobility-driven matrices was $0.75$ among
adults and $0.67$ among seniors (Fig.~\ref{fig:matrix_structure}d). By contrast,
matrix elements involving children or adolescents were close to the mobility-driven scale, with $b=0.99$ and $r=0.90$
(Fig.~\ref{fig:matrix_structure}b). The corresponding age-specific
contact-intensity ratios were $0.92$ for children and $1.01$ for adolescents.
GPT-4o mini therefore differed from the mobility-driven matrices primarily through
lower adult and senior contact volumes rather than through a uniform reduction
across all age groups.

Gemini 2.5 Flash generated higher values in the adult--senior block
($b=1.14$, $r=0.94$; Fig.~\ref{fig:matrix_structure}a), but this elevation
was not distributed uniformly between adults and seniors. Median contact
intensity was $1.21$ times the mobility-driven level among adults but $0.89$ times
the mobility-driven level among seniors (Fig.~\ref{fig:matrix_structure}d). Elements
involving children or adolescents were lower overall, with $b=0.78$ and
$r=0.87$ (Fig.~\ref{fig:matrix_structure}b). This reduction was concentrated
among adolescents, for whom the median ratio was $0.84$, compared with $0.96$
among children. Gemini 2.5 Flash therefore generated an age allocation
characterized by comparatively high adult contact intensity and reduced
adolescent and senior contact intensity.

Grok 3 mini generated lower contact values in both element sets. Its
adult--senior slope was $b=0.86$, with $r=0.90$
(Fig.~\ref{fig:matrix_structure}a), and its young-related slope was
$b=0.75$, with $r=0.82$ (Fig.~\ref{fig:matrix_structure}b). Median
contact-intensity ratios were below one in all four age groups: $0.82$ among
children, $0.79$ among adolescents, $0.89$ among adults, and $0.71$ among
seniors (Fig.~\ref{fig:matrix_structure}d). Grok 3 mini therefore differed
from the mobility-driven matrices primarily through broadly lower contact intensity,
although the magnitude of the reduction varied across ages.

The within-group contact fractions provided an additional comparison of mixing
structure that was independent of overall contact volume
(Fig.~\ref{fig:matrix_structure}c). GPT-4o mini produced comparatively high
within-group fractions among children and adolescents, whereas Gemini 2.5 Flash
and Grok 3 mini produced lower values in these younger groups. Differences among
adults and seniors were smaller, with all time-varying configurations retaining
comparatively strong adult assortativity and weaker senior assortativity.

Taken together, GPT-4o mini closely reproduced the scale of matrix elements
involving younger groups but reduced adult and senior contact intensity.
Gemini 2.5 Flash redistributed contact intensity toward adults and away from
adolescents and seniors, together with lower within-group mixing among younger
groups. Grok 3 mini generated lower contact intensity across all four age
groups and showed larger departures for elements involving younger ages. These
structural patterns provide additional context for the age-specific
hospitalization biases observed in the retrospective reconstruction.

\subsection*{Validation under the factual policy scenario}

Counterfactual outcomes cannot be validated directly because the alternative
policy scenarios were not observed. We therefore evaluated the preceding step of
the framework: whether it could reproduce the subsequent epidemic trajectory
when supplied with the policy sequence that was actually implemented, while
remaining blinded to future hospital admissions. We refer to this as the
policy-conditioned configuration.

The policy-conditioned analysis used the same 13 rolling origins, the weeks beginning 14 September to 7 December 2020, and the same four-week horizon as the real-time forecasting analysis. It
differed from the real-time configuration only in the policy block and its
duration counter. In the real-time forecasts, the policy levels observed at the
origin were held throughout the horizon. In the policy-conditioned
configuration, the policy block and duration counter were instead taken from
the target week, thereby supplying the control measures that were subsequently
implemented.

All other components were held identical to the real-time forecasts, including the participant panel, prompt template, school calendar, fitted transmission parameters, correcting-factor smoothing rule, and ensemble size. The epidemic block also followed the same real-time information constraint. 
Hospital admissions reported up to the origin were
available, and admissions for earlier weeks within the projected horizon were
taken from the model's own median predictions. No hospital admission observed
after the origin was used to generate the matrices or simulate the projection.

\begin{figure}[p]
\centering
\includegraphics[width=\textwidth]{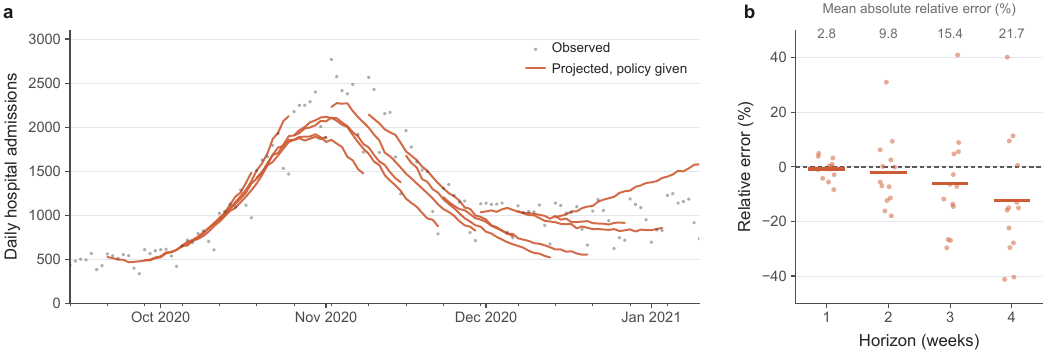}
\caption{\textbf{Validation under the factual policy scenario across 13 rolling
origins.}
(\textbf{a}) Observed daily hospital admissions and four-week projections when
the policy levels subsequently implemented are supplied for each target week,
while future admissions remain unavailable.
(\textbf{b}) Relative error by forecast horizon. Horizontal bars denote means,
and annotations report mean absolute relative error.}
\label{fig:val}
\end{figure}

Given the factual policy scenario, the framework reproduced the principal
turning points of the autumn 2020 wave across the 13 origins
(Fig.~\ref{fig:val}). Projections issued during the rise continued upward, those
spanning the imposition of the second national lockdown turned downward, and
those issued during the decline preserved the falling trajectory.

Mean absolute relative error increased with forecast horizon, from $2.8\%$ at
one week to $21.7\%$ at four weeks. Mean 95\% interval coverage across all
origin--horizon targets was $96\%$. Signed errors were centered close to zero at
one week and became increasingly negative with horizon, reaching a mean of
$-12.2\%$ at four weeks. The framework therefore tended to overstate the
reduction in admissions produced by the factual policy scenario at the longest
horizon.

The factual-policy projections used in the two counterfactual experiments form
part of this broader rolling-origin evaluation. At the branch preceding the
second national lockdown, the factual projection underestimated cumulative
admissions over the four-week window by $9.7\%$. At the branch preceding the
staged relaxation, the corresponding bias was $-2.9\%$. These projections
provide the factual benchmark against which the alternative policy scenarios in
the main text are compared. They do not establish that the counterfactual
outcomes are correct, but show that the framework responded coherently when the
subsequently observed policy sequence was supplied without access to future
hospitalization outcomes.

\subsection*{Sensitivity to historical information leakage}

Retrospective evaluation under historical COVID-19 conditions introduces a
potential source of information leakage: a LLM may recognize the
historical episode from identifying cues in the prompt and draw on information
acquired during pretraining. The main analysis limits such recognition by
withholding calendar dates and describing the pathogen generically. We further
tested the sensitivity of the behavioral layer to historical identifiability by
perturbing these cues in both directions while holding the behavioral task and
downstream modeling pipeline fixed. All analyses in this section used GPT-4o
mini.

We compared three prompt-information conditions. The baseline condition was the masked prompt used throughout the main analysis. The deeper-masking condition removed the country name and replaced numerical epidemic quantities with prespecified qualitative descriptions of epidemic
level and week-to-week trend. The full-disclosure condition instead identified the pathogen as COVID-19 caused by SARS-CoV-2 and supplied the calendar date corresponding to each target week. All other inputs and procedures were unchanged across conditions.
\paragraph{Prompt modifications.}
Under deeper masking, references to France in the disease background, policy
header, and participant description were replaced by geographically generic
wording. Numerical epidemic information was converted to qualitative categories
using thresholds specified before the sensitivity analysis. Weekly counts were
normalized per million population. New cases were categorized using cut points
of $50$, $300$, $1500$, and $5000$ per million, and new deaths using $1$, $5$,
$20$, and $60$ per million, defining very low, low, moderate, high, and very
high levels. For prospective prompts based on hospital admissions, the
corresponding cut points were $15$, $50$, $120$, and $250$ per million.
Cumulative deaths and admissions were categorized as limited, substantial,
large, or very large using cut points of $15$, $150$, and $500$ per million
for deaths and $300$, $1500$, and $4500$ per million for admissions. Weekly
trend was determined from the ratio of the current to previous value and
classified as rising steeply ($\geq1.5$), rising ($1.15$ to $<1.5$), roughly
stable ($0.87$ to $<1.15$), declining ($0.67$ to $<0.87$), or declining
steeply ($<0.67$). These thresholds were fixed and were not estimated from
contact or hospitalization outcomes. Under full disclosure, the task explicitly identified the epidemic as COVID-19
in France, described the pathogen as SARS-CoV-2, and supplied the calendar week.
The numerical epidemic information and all other contextual fields were
otherwise identical to the baseline condition. 

\paragraph{Worked examples.}

The following prompts reproduce the complete messages used under the deeper-masking and full-disclosure conditions for the same participant and target week as the baseline prompt shown in the LLM prompts and queries subsection.

\medskip
\noindent\textbf{Deeper-masking condition}
\par\smallskip

\begin{systemprompt}
You are simulating how one real person behaved during an epidemic.

Background: a febrile respiratory infectious disease is spreading in the country where you live. It transmits through close human contact and can cause serious illness. The government responds with measures such as lockdowns, curfews, school and workplace closures, and limits on gatherings to reduce transmission.

The six contact settings, each corresponding to a physical location of contact:

- home: contacts at home, including the people you live with and any visitors
- work: contacts at the workplace
- school: contacts at school, college, or university
- transport: contacts on public transport
- leisure: contacts during leisure activities such as bars, restaurants, sports, cultural venues
- otherplace: contacts at other locations such as shops, services, and other public places

You will be given one person's age and their actual contact diary from a national survey carried out before the epidemic, on an ordinary working day. The diary lists every single person they had face-to-face or physical contact with on that day, with that person's age and where the contact took place. These are real contacts reported by a real respondent, not estimates.

Role-play that person under the conditions described below.

First work out, from the diary alone, what this person's ordinary day looks like. Are they at school, in work, retired, or at home? If they work, what kind of work is it, and could it be done from home? Who do they live with? The number of contacts they have in each setting, and the ages of those contacts, tell you most of this.

Then go through the diary one contact at a time and decide, for each one, whether that particular contact would still take place on a comparable day in the current week.

Decide each contact on its own merits. Some considerations push towards giving a contact up:

- the law currently forbids or restricts the activity
- the setting is crowded, indoors, or otherwise risky
- the epidemic is severe or getting worse this week

Other considerations push towards keeping a contact:

- the activity is essential or unavoidable for this person
- giving it up would cost them their income or their job
- someone depends on them for care
- restrictions have been in force for many weeks now, and people have grown tired of them
- compliance is never perfect, and some restricted contact always continues

Weigh both sides against this specific person and this specific contact. Model realized behaviour, not perfect compliance, and do not assume the person obeys every rule.

One thing to keep in mind. Contacts at home include the people you live with, and you keep seeing them every day even under a strict lockdown.

Report the numbers of the contacts that would still take place. Any number you leave out is treated as a contact that no longer takes place. If none of them would still take place, report an empty list.

Respond ONLY with a JSON object, no other text:

{"still_happen": [list of contact numbers]}

============================================================

Current control policies in your country:

- Schools: some categories of schools required to close
- Workplaces: all but essential workplaces required to close
- Public events: public events required to be cancelled
- Gatherings: restrictions on gatherings of 10 people or fewer
- Stay-at-home: required not to leave home except for essential trips, exercise, and grocery shopping
- Public transport: no public transport measures
- Internal movement: internal movement restrictions in place

These measures came into force this week.

Epidemic situation this week, described in broad terms:

- New confirmed cases this week: a high level for the size of the population, roughly stable compared with last week
- New deaths this week: a very high level for the size of the population, rising steeply compared with last week
- Cumulative deaths since the start of the epidemic: very large relative to the population

School calendar: schools are in their normal teaching term.

Control measures of some kind have been in force for 33 weeks in total.
\end{systemprompt}

\begin{userprompt}
You are a 45 year old person.

Your own contact diary from before the epidemic, on an ordinary working day. You had face-to-face or physical contact with these 8 people:
1. a person aged about 25 to 30, met at a shop, a service, or another public place
2. a person aged about 25 to 30, met at work
3. a person aged about 30 to 35, met at work
4. a person aged about 40 to 45, met at a shop, a service, or another public place
5. a person aged about 50 to 55, met on public transport
6. a person aged about 25 to 30, met during a leisure activity
7. a person aged about 40 to 45, met at home
8. a person aged about 40 to 45, met at home

Which of these contacts would still take place on a comparable day this week? Report their numbers.
\end{userprompt}

\medskip
\noindent\textbf{Full-disclosure condition}
\par\smallskip

\begin{systemprompt}
You are simulating how one real person behaved during the COVID-19 epidemic in France.

Background: COVID-19, a disease caused by the SARS-CoV-2 coronavirus, is spreading in France. It transmits through close human contact and can cause serious illness. The government responds with measures such as lockdowns, curfews, school and workplace closures, and limits on gatherings to reduce transmission.

The six contact settings, each corresponding to a physical location of contact:

- home: contacts at home, including the people you live with and any visitors
- work: contacts at the workplace
- school: contacts at school, college, or university
- transport: contacts on public transport
- leisure: contacts during leisure activities such as bars, restaurants, sports, cultural venues
- otherplace: contacts at other locations such as shops, services, and other public places

You will be given one person's age and their actual contact diary from a national survey carried out before the epidemic, on an ordinary working day. The diary lists every single person they had face-to-face or physical contact with on that day, with that person's age and where the contact took place. These are real contacts reported by a real respondent, not estimates.

Role-play that person under the conditions described below.

First work out, from the diary alone, what this person's ordinary day looks like. Are they at school, in work, retired, or at home? If they work, what kind of work is it, and could it be done from home? Who do they live with? The number of contacts they have in each setting, and the ages of those contacts, tell you most of this.

Then go through the diary one contact at a time and decide, for each one, whether that particular contact would still take place on a comparable day in the current week.

Decide each contact on its own merits. Some considerations push towards giving a contact up:

- the law currently forbids or restricts the activity
- the setting is crowded, indoors, or otherwise risky
- the epidemic is severe or getting worse this week

Other considerations push towards keeping a contact:

- the activity is essential or unavoidable for this person
- giving it up would cost them their income or their job
- someone depends on them for care
- restrictions have been in force for many weeks now, and people have grown tired of them
- compliance is never perfect, and some restricted contact always continues

Weigh both sides against this specific person and this specific contact. Model realized behaviour, not perfect compliance, and do not assume the person obeys every rule.

One thing to keep in mind. Contacts at home include the people you live with, and you keep seeing them every day even under a strict lockdown.

Report the numbers of the contacts that would still take place. Any number you leave out is treated as a contact that no longer takes place. If none of them would still take place, report an empty list.

Respond ONLY with a JSON object, no other text:

{"still_happen": [list of contact numbers]}

============================================================

The current week is the week beginning Monday 2 November 2020.

Current control policies in France:

- Schools: some categories of schools required to close
- Workplaces: all but essential workplaces required to close
- Public events: public events required to be cancelled
- Gatherings: restrictions on gatherings of 10 people or fewer
- Stay-at-home: required not to leave home except for essential trips, exercise, and grocery shopping
- Public transport: no public transport measures
- Internal movement: internal movement restrictions in place

These measures came into force this week.

Epidemic situation this week:

- New confirmed cases this week: 303,116, last week: 332,505
- Cumulative confirmed cases: 1,643,952
- New deaths this week: 4,794, last week: 3,154
- Cumulative deaths: 33,509

School calendar: schools are in their normal teaching term.

Control measures of some kind have been in force for 33 weeks in total.
\end{systemprompt}

\begin{userprompt}
You are a 45 year old person in France.

Your own contact diary from before the epidemic, on an ordinary working day. You had face-to-face or physical contact with these 8 people:
1. a person aged about 25 to 30, met at a shop, a service, or another public place
2. a person aged about 25 to 30, met at work
3. a person aged about 30 to 35, met at work
4. a person aged about 40 to 45, met at a shop, a service, or another public place
5. a person aged about 50 to 55, met on public transport
6. a person aged about 25 to 30, met during a leisure activity
7. a person aged about 40 to 45, met at home
8. a person aged about 40 to 45, met at home

Which of these contacts would still take place on a comparable day this week? Report their numbers.
\end{userprompt}

\begin{figure}[p]
\centering
\includegraphics[
  width=\textwidth,
  height=0.85\textheight,
  keepaspectratio
]{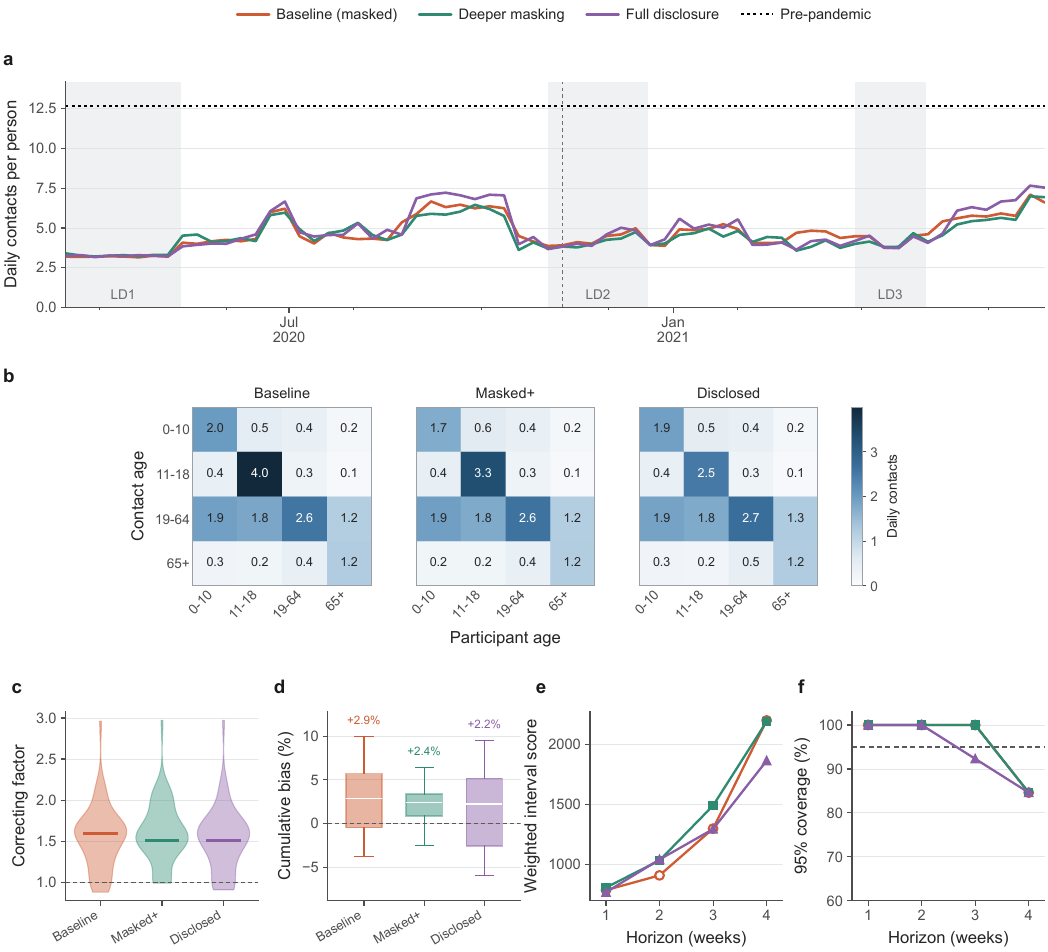}
\caption{\textbf{Sensitivity of generated contact patterns and epidemiological
performance to epidemic-context information.}
GPT-4o mini was evaluated under three information conditions: the masked
baseline used in the main analysis, deeper masking that additionally removes
the country identity and replaces numerical epidemic counts with qualitative
descriptions, and full disclosure that explicitly identifies COVID-19, France,
and the calendar week.
\textbf{(a)} Weekly aggregate daily contacts per person generated under the
three conditions. The dotted horizontal line indicates the pre-pandemic contact
level; shaded intervals denote the three national lockdowns.
\textbf{(b)} Age-structured contact matrices in the week beginning 9 November 2020, under baseline, deeper masking, and full disclosure.
\textbf{(c)} Distribution of calibration correcting factors across
retrospective calibration windows.
\textbf{(d)} Cumulative hospitalization bias in the retrospective evaluation;
labels report the corresponding mean bias.
\textbf{(e)} Weighted interval score by forecast horizon.
\textbf{(f)} Empirical coverage of the $95\%$ prediction intervals by forecast
horizon; the dashed horizontal line marks nominal $95\%$ coverage.
}
\label{fig:leakage}
\end{figure}

The generated behavioral trajectories were stable across the three information
conditions (Fig.~\ref{fig:leakage}a,b). Across 68 matched weeks, aggregate
contact intensity under deeper masking and full disclosure remained strongly
correlated with the baseline trajectory ($r=0.92$ and $r=0.95$,
respectively). Mean absolute relative differences from baseline were $6.3\%$
and $7.0\%$, whereas mean signed differences were small and opposite in
direction ($-2.2\%$ and $+2.1\%$). During the second national lockdown, mean absolute deviations from baseline were $4.4\%$ under
deeper masking and $3.8\%$ under full disclosure. The age structure of the
generated matrices was similarly preserved. Across matched weeks, element-wise
comparisons with the baseline yielded regression slopes through the origin of
$0.89$ under deeper masking and $1.05$ under full disclosure, with Pearson
correlations of $0.98$ for both conditions. Median age-specific contact-intensity
ratios relative to baseline ranged from $0.91$ to $1.04$. The matrices for the week beginning 9 November 2020 provide a representative illustration of this stability in age-specific mixing
structure (Fig.~\ref{fig:leakage}b).

Differences between information conditions also remained limited after
propagation through the epidemiological model
(Fig.~\ref{fig:leakage}c--f). The distributions of fitted correcting factors
were highly similar across the three conditions, with median values close to
$1.5$ in each case (Fig.~\ref{fig:leakage}c). Retrospective cumulative
hospitalization bias was $+2.9\%$ under the baseline, $+2.4\%$ under deeper
masking, and $+2.2\%$ under full disclosure
(Fig.~\ref{fig:leakage}d). Rolling four-week forecasts were likewise similar
across conditions. Mean weighted interval score was $1298$ under the
baseline, $1380$ under deeper masking, and $1241$ under full disclosure
(Fig.~\ref{fig:leakage}e), corresponding to WIS ratios relative to baseline of
$1.06$ and $0.96$ for the two alternative conditions. Coverage of the $95\%$
prediction intervals was $96.2\%$ under the baseline and deeper-masking
conditions and $94.2\%$ under full disclosure
(Fig.~\ref{fig:leakage}f).

The two perturbations therefore provide complementary evidence about dependence
on historical identity cues. Removing the country name and absolute epidemic
counts did not materially alter either the generated contact structure or its
downstream epidemiological performance. Conversely, explicitly identifying
COVID-19, France, and the historical calendar week did not produce a systematic
improvement over the masked baseline. Together, these results show that the
main behavioral and epidemiological findings are insensitive to substantial
changes in the historical identifiability of the prompt. 

\clearpage

\begin{table}[p]
\centering\small
\caption{\textbf{Fields of the system message that differ between settings}. All other content is identical throughout.}
\label{tab:settings}
\vspace{0.5em}
\begin{tabular}{p{3.2cm} p{3.2cm} p{3.5cm} p{4cm}}
\hline
Setting & Policy levels at week $t$ & Duration counter & Epidemic block \\
\hline
Retrospective & observed at $t$ & observed run ending at $t$ & cases and deaths at $t$ \\
Forecast from origin $o$ & observed at $o$ & run at $o$, extended by $t-o$ & admissions to $t-1$, simulated after $o$ \\
Policy-conditioned & observed at $t$ & observed run ending at $t$ & admissions to $t-1$, simulated after $o$ \\
Counterfactual & scenario table at $t$ & run on the scenario path & admissions to $t-1$, simulated after $o$ \\
\hline
\end{tabular}
\end{table}

\begin{table}[p]
\centering\small
\caption{\textbf{Natural-history and clinical parameters of the transmission model.} Values
apply to both strains except where a strain is named. The asymptomatic
probability is the same for all ages and both strains.}
\label{tab:nat}
\vspace{0.5em}
\begin{tabular}{p{5.4cm} l p{2.4cm}}
\hline
Parameter & Value & Source \\
\hline
Latent period, $1/\sigma$ & 3.7 days & \cite{lauer2020incubation} \\
Presymptomatic period, $1/\theta$ & 1.5 days & \cite{ferretti2020quantifying} \\
Infectious period, $1/\gamma$ & 2.0 days & \cite{cereda2021early} \\
Waiting time for hospitalization, $1/\omega$ & 3.0 days & \cite{boelle2020trajectories,pellis2021challenges} \\
Time in hospital, $1/\kappa$ & 14.0 days & \cite{boelle2020trajectories} \\
Probability asymptomatic, $p_{as}$ & 0.4 & \cite{lavezzo2020suppression} \\
Probability of pauci-symptomatic, $p_{ps}$ (ch, ado / adu, sen) & 0.4 / 0.12 & \cite{riccardo2020epidemiological,davies2020age} \\
Probability of severe symptoms, $p_{ss}$, Wuhan-like (ch / ado / adu / sen) & 0.002 / 0.001 / 0.011 / 0.096 & \cite{lapidus2021severe} \\
Relative infectiousness of $I_p$, $I_{as}$, $I_{ps}$ with respect to $I_{ms}$ and $I_{ss}$, $r$ (children / adolescents, adults, seniors) & 0.25 / 0.55 & \cite{li2020substantial} \\
Susceptibility, Wuhan-like (children, adolescents / adults, seniors) & 0.7 / 1.0 & \cite{davies2020age,hu2021infectivity,franco2022inferring,viner2021susceptibility} \\
Susceptibility, Alpha (all) & 1.0 & \cite{didomenico2026natcommun} \\
Alpha transmission advantage, $\eta$ & 1.59 & \cite{gaymard2021alpha} \\
Alpha hospitalization multiplier & 1.64 & \cite{paredes2022hospitalization} \\
\hline
\end{tabular}
\end{table}
 
\begin{table}[p]
\centering\small
\caption{\textbf{Vaccine effectiveness by dose level, applied equally to both strains.}The values are for the Pfizer
vaccine. Dose roll-out follows the administered-dose record \cite{spf_vacsi}, and the
maximum coverage among adults and seniors is 0.99.}
\label{tab:vax}
\vspace{0.5em}
\begin{tabular}{p{5cm} c c p{2.4cm}}
\hline
Vaccine effect & One dose & Two doses & Source \\
\hline
Against infection & 0.60 & 0.95 & \cite{dagan2021bnt162b2,haas2021impact,sheikh2021delta} \\
Against onward transmission given infection & 0.15 & 0.68 & \cite{eyre2022effect} \\
Against symptomatic disease & 0.70 & 0.97 & \cite{dagan2021bnt162b2,haas2021impact} \\
Against severe disease & 0.80 & 0.975 & \cite{dagan2021bnt162b2,haas2021impact} \\
\hline
\end{tabular}
\end{table}

\begin{table}[p]
\centering\small
\caption{\textbf{Candidate branch points, ranked by the simulated impact of the policy
change observed over the window.} Origins and targets are identified by the
Monday of the corresponding week; the target is four weeks after the origin.
$\Delta$SI is the change in the tracker's average stringency index between the
origin and the target week.}
\label{tab:cf_cand}
\vspace{0.5em}
\begin{tabular}{l l r r p{5.6cm}}
\hline
Origin & Target & $\Delta$SI & Impact & Indicators that moved \\
\hline
26 Oct 2020 & 23 Nov 2020 & $15.6$ & $+131\%$ & schools, workplaces, public events, internal movement \\
12 Oct 2020 & 9 Nov 2020 & $33.1$ & $+120\%$ & schools, workplaces, public events, stay-at-home, internal movement \\
19 Oct 2020 & 16 Nov 2020 & $29.2$ & $+46\%$ & schools, workplaces, public events, internal movement \\
23 Nov 2020 & 21 Dec 2020 & $-13.8$ & $-31\%$ & workplaces, internal movement \\
14 Sep 2020 & 12 Oct 2020 & $-3.0$ & $+25\%$ & schools, internal movement \\
7 Dec 2020 & 4 Jan 2021 & $-11.1$ & $-15\%$ & internal movement \\
5 Oct 2020 & 2 Nov 2020 & $30.7$ & $+12\%$ & schools, workplaces, stay-at-home, internal movement \\
28 Sep 2020 & 26 Oct 2020 & $12.5$ & $+5\%$ & public events, stay-at-home, internal movement \\
16 Nov 2020 & 14 Dec 2020 & $-13.2$ & $-4\%$ & workplaces, internal movement \\
2 Nov 2020 & 30 Nov 2020 & $-3.7$ & $-4\%$ & workplaces \\
21 Sep 2020 & 19 Oct 2020 & $1.7$ & $+3\%$ & stay-at-home, internal movement \\
9 Nov 2020 & 7 Dec 2020 & $-3.7$ & $-2\%$ & workplaces \\
30 Nov 2020 & 28 Dec 2020 & $-11.1$ & $-1\%$ & internal movement \\
\hline
\end{tabular}
\end{table}

\begin{table}[ht]
\centering\small
\caption{\textbf{Scenario level tables.} Each cell lists the seven tracker levels for that
week, in the order C1--C7: schools, workplaces, public events,
gatherings, public transport, stay-at-home, and internal movement. Column
headers give the Monday of each projected week; the first projected week
begins one week after the branch.}
\label{tab:cf_scen}
\vspace{0.5em}
\begin{tabular}{l l l l l}
\hline
\multicolumn{5}{l}{\textbf{Lockdown branch}, levels in force at the branch $1|2|1|4|0|2|0$} \\
Scenario & 2 Nov & 9 Nov & 16 Nov & 23 Nov \\
\hline
Factual & $2|3|2|4|0|2|2$ & $2|3|2|4|0|2|2$ & $2|3|2|4|0|2|2$ & $2|3|2|4|0|2|2$ \\
No change & $1|2|1|4|0|2|0$ & $1|2|1|4|0|2|0$ & $1|2|1|4|0|2|0$ & $1|2|1|4|0|2|0$ \\
Delayed 1 wk & $1|2|1|4|0|2|0$ & $2|3|2|4|0|2|2$ & $2|3|2|4|0|2|2$ & $2|3|2|4|0|2|2$ \\
Delayed 2 wk & $1|2|1|4|0|2|0$ & $1|2|1|4|0|2|0$ & $2|3|2|4|0|2|2$ & $2|3|2|4|0|2|2$ \\
Maximum & $3|3|2|4|2|3|2$ & $3|3|2|4|2|3|2$ & $3|3|2|4|2|3|2$ & $3|3|2|4|2|3|2$ \\
Schools closed & $3|3|2|4|0|2|2$ & $3|3|2|4|0|2|2$ & $3|3|2|4|0|2|2$ & $3|3|2|4|0|2|2$ \\
\hline
\multicolumn{5}{l}{\textbf{Relaxation branch}, levels in force at the branch $2|3|2|4|0|2|2$} \\
Scenario & 30 Nov & 7 Dec & 14 Dec & 21 Dec \\
\hline
Factual & $2|2|2|4|0|2|2$ & $2|2|2|4|0|2|2$ & $2|2|2|4|0|2|0$ & $2|2|2|4|0|2|0$ \\
No change & $2|3|2|4|0|2|2$ & $2|3|2|4|0|2|2$ & $2|3|2|4|0|2|2$ & $2|3|2|4|0|2|2$ \\
Delayed 1 wk & $2|3|2|4|0|2|2$ & $2|2|2|4|0|2|2$ & $2|2|2|4|0|2|2$ & $2|2|2|4|0|2|0$ \\
Delayed 2 wk & $2|3|2|4|0|2|2$ & $2|3|2|4|0|2|2$ & $2|2|2|4|0|2|2$ & $2|2|2|4|0|2|2$ \\
Maximum & $3|3|2|4|2|3|2$ & $3|3|2|4|2|3|2$ & $3|3|2|4|2|3|2$ & $3|3|2|4|2|3|2$ \\
Schools closed & $3|2|2|4|0|2|2$ & $3|2|2|4|0|2|2$ & $3|2|2|4|0|2|0$ & $3|2|2|4|0|2|0$ \\
\hline
\end{tabular}
\end{table}

\end{document}